\documentstyle[epsfig,epsf,floats]{jfm}

\newcommand{\aleq}{\mbox{\ 
\raisebox{-.9ex}{$\stackrel{\textstyle<}{\sim}$}\ }}
\newcommand{\ageq}{\mbox{\
\raisebox{-.9ex}{$\stackrel{\textstyle >}{\sim}$}\ }}

\def\la{{\langle}}
\def\ra{{\rangle}}
\def\om{{\omega}}
\def\begineq{\begin{equation}}
\def\endeq{\end{equation}}
\def\etal{\mbox{\it et al.\ }}

\begin{document}

\title[Rayleigh--Plesset dynamics for sonoluminescing bubbles]{
Analysis of Rayleigh--Plesset dynamics for sonoluminescing
bubbles}
\author[S.~Hilgenfeldt, M.~P.~Brenner, S.~Grossmann, and
D.~Lohse]
{\\[\affilskip]
S\ls A\ls S\ls C\ls H\ls A\ns
H\ls I\ls L\ls G\ls E\ls N\ls F\ls E\ls L\ls D\ls T$^1$,\ns
M\ls I\ls C\ls H\ls A\ls E\ls L\ns
P.\ns B\ls R\ls E\ls N\ls N\ls E\ls R$^2$,\ns \\[\affilskip]
S\ls I\ls E\ls G\ls F\ls R\ls I\ls E\ls D\ns
G\ls R\ls O\ls S\ls S\ls M\ls A\ls N\ls N$^1$,\ns \and
D\ls E\ls T\ls L\ls E\ls F\ns
L\ls O\ls H\ls S\ls E$^1$}
\affiliation{$^1$
Fachbereich Physik der Universit\"at Marburg,
Renthof 6, 35032 Marburg, Germany \\[\affilskip]
$^2$ Department of Mathematics, MIT, Cambridge, MA 02139}  

\date{22 February 1997 and in revised form 21 January 1998}

\maketitle
\begin{abstract}
Recent work on single bubble sonoluminescence (SBSL) has shown
that many features of this phenomenon, especially the dependence of   
SBSL intensity and stability on experimental parameters, can be explained
within a {\em hydrodynamic approach}. More specifically, many important
properties can already be
derived from an analysis of {\em bubble wall dynamics}. This dynamics is
conveniently described by the Rayleigh--Plesset (RP) equation. In this work
we derive analytical approximations for RP dynamics
and subsequent analytical laws for parameter dependences.
These results include
(i) an expression for the onset threshold of SL,
(ii) an analytical explanation of the transition from diffusively unstable to
stable equilibria for the bubble ambient radius (unstable and stable
sonoluminescence), and (iii) a detailed understanding of
the resonance structure of the RP equation.
It is found that the threshold for SL emission is shifted to larger
bubble radii and larger driving pressures if surface tension is enlarged,
whereas even a considerable change in liquid viscosity leaves this threshold
virtually unaltered. As an enhanced viscosity stabilizes the bubbles against
surface oscillations, we conclude that the ideal liquid for violently
collapsing, surface stable SL bubbles should have small surface tension and
large viscosity, although too large viscosity ($\eta_l\ageq40\eta_{\em water}$)
will again preclude collapses.
\end{abstract}

%----------------------------------------------------------------------
\section{Introduction}\label{secintro}

\subsection{Sonoluminescence}\label{secsl}

The analysis of the dynamics of a small bubble or cavity in a fluid dates
back to the work of Lord \cite{ray17} at the beginning of this century.
A large number of publications followed in subsequent
decades, including the studies of oscillating bubbles by
Plesset (1949,\,1954), \cite{ell70}, Flynn (1975a,\,1975b),
\cite{lau76,pro77,ple77}, and others.
In recent years,
a renascence of bubble dynamics has occurred initiated by the
discovery of single bubble sonoluminescence (SBSL) by
\cite{gai90}, see also \cite{gai92}.

SBSL is an intriguing phenomenon: A single gas bubble
of only a few $\mu m$ size, levitated in water by an acoustic standing
wave, emits light pulses so intense as to be visible to the naked eye.
The standing ultrasound wave of the driving
keeps the bubble in position at a pressure antinode and, at the same time,
drives its oscillations.
The experiments of Putterman's group
(Barber \& Putterman 1991; Barber \etal (1994,\,1995); Hiller \etal 1994;
L\"ofstedt, Barber \& Putterman 1993; L\"ofstedt \etal 1995; Weninger,
Putterman \& Barber 1996) and others
have revealed a multitude of interesting facts about SBSL: the width of the
light pulse is small (Barber \& Putterman 1991 give $50\,ps$ as
upper threshold, Moran \etal 1995 $10\,ps$ -- recent measurements by Gompf
\etal 1997 report 100-300\,$ps$, depending on the forcing pressure and
gas concentration in the liquid),
the spectrum shows no features
such as lines (Hiller, Putterman \& Barber 1992; Matula \etal 1995).
While the exact mechanism of light emission is
still an open issue, almost all suggested theories -- see e.g.\ 
L\"ofstedt \etal (1993), Hiller \etal (1992), \cite{fli89}, \cite{wu93},
\cite{fro94}, Moss \etal (1994),
Bernstein \& Zakin (1995), \cite{mos97} --
agree that temperatures of at least $10^4$-$10^5\,K$ are reached during bubble
collapse. This, together with the light intensity, clearly shows that SBSL
relies on 
an extraordinarily powerful energy focusing process.

In our previous publications \cite{bre95}, Brenner \etal (1996a,\ 1996b),
\cite{hil96}, \cite{bre96b},\linebreak \cite{loh97}, and \cite{loh97b} we
calculated phase diagrams for bubbles and
have focused on the identification of
parameter regimes where SBSL occurs.
As a scan of the whole multi-dimensional parameter space is by far too
expensive for full numerical simulations of the underlying fundamental
equations (i.e., Navier--Stokes and advection-diffusion PDEs), it is necessary
to introduce approximations.
The necessary conditions for SL to occur could be calculated from the
dynamics $R(t)$ of the bubble wall, which is -- apart from a tiny interval
around the bubble collapse -- very well described by
the Rayleigh--Plesset (RP) equation. We call this approach the
{\em RP-SL bubble approach}.

The key parameters in an SL experiment are the ambient bubble radius $R_0$
(radius under normal conditions of $1.013\times 10^5\,{\mbox{\it Pa}}=1\,atm$
and
$20^\circ$C),
the driving pressure amplitude
$P_a$, and the gas concentration in the water surrounding the bubble
$p_\infty/P_0$,
measured by its partial pressure divided by the ambient pressure.
Note that $R_0$ is not at the experimenter's disposal, but adjusts itself
by gas diffusion on a slow time scale of seconds. Its size can, however,
be measured in experiment, e.g., by Mie scattering techniques as in
\cite{bar95}
or by direct microscopic imaging, see \cite{tia96,hol96}. On time scales
much smaller than those of diffusive processes, e.g.\ for one period of
driving, $R_0$ may be regarded as a constant to high accuracy.

In Hilgenfeldt \etal (1996) we found that the
$P_a/P_0 - p_\infty/P_0$ state space is divided into
regions where (diffusively) {\em stable SL}, {\em unstable SL} or {\em no SL}
are to be expected, in excellent agreement with
experimental findings. These results will now be briefly presented in the
following subsection.

\subsection{Stability requirements}\label{secstab}

Stable sonoluminescence is characterized by light
emission in each period of driving at precisely the same oscillation phase
and precisely the same brightness for millions (and sometimes billions)
of cycles. We found that it occurs in a tiny section of the whole
parameter space only, and that the calculated domain agrees very well
with experimental findings, cf.\ Hilgenfeldt \etal (1996), \cite{loh97}.
Its boundaries are set by certain
dynamical and stability conditions imposed upon the oscillating bubble
(Brenner \etal 1995, Brenner \etal 1996a, Hilgenfeldt \etal 1996):
(i) The bubble wall velocity during collapse must reach the speed of sound
in the gas $c_{g}$ to ensure sufficient
energy transfer from the liquid to the gas.
(ii) The bubble must be stable towards
non-spherical oscillations of its surface which lead
to fragmentation.
Bubble fragments have meanwhile been experimentally
observed by J.~Holzfuss (private communication, 1997).
(iii) The bubble must be stable towards
diffusive processes, i.e., it must not dissolve or grow by rectified
diffusion; diffusively growing bubbles show {\em unstable SL}.  
A further requirement of (iv) chemical stability becomes important when the
bubble contains molecular gases which are able to dissociate and recombine
with liquid molecules (Brenner \etal 1996, Lohse \etal 1997).
E.g., the differences in the parameter regimes of SL in air bubbles vs.\
SL in noble gas bubbles can consistently be accounted for by dissociation
of N$_2$ and O$_2$ in an air bubble;
these molecular constituents of air are burned,
leaving only inert gases in the bubble (the experimental work of Holt \&
Gaitan 1996 supports this model). 
We therefore
restrict ourselves for simplicity to the case of a bubble filled with argon.
An extension to reactive gas mixtures as analysed in \cite{loh97} is
straightforward.
Also, we specify the liquid in which the bubble oscillates to
be water, as in most SBSL experiments.

%caption1
\begin{figure}[htb]
\setlength{\unitlength}{1.0cm}
\begin{picture}(12,11)
\put(0.5,0.5){\psfig{figure=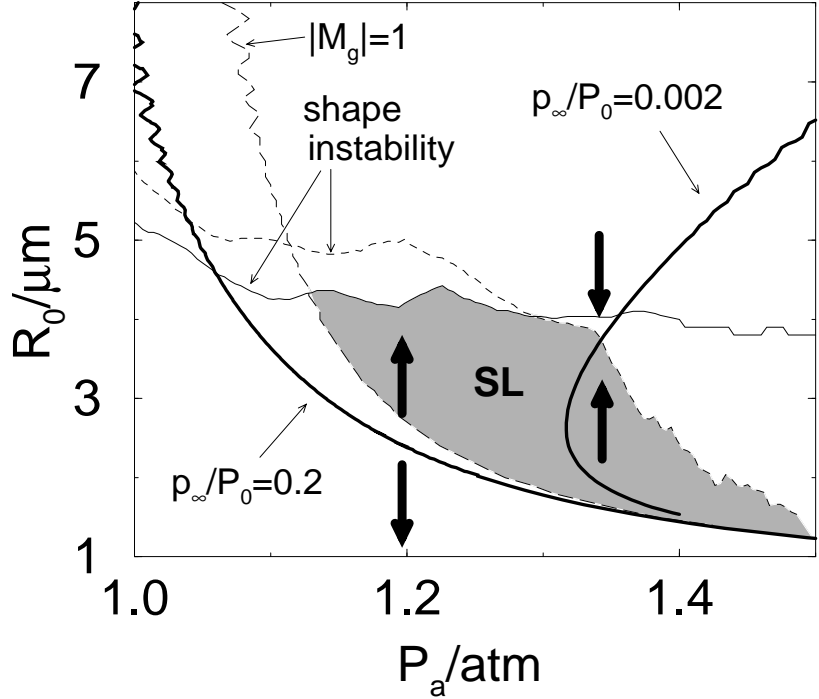,width=13cm,angle=-90.}}
\end{picture}
\caption[]{Stability conditions for a bubble in $P_a$--$R_0$ parameter space.
Bubbles above the line $|M_{g}|=1$ fulfill the energy focusing condition
(\ref{mach}).
Bubbles below the shape instability lines are stable towards non-spherical
surface oscillations. The solid line represents the (long time scale)
parametric instability, the dashed line short time scale shape instabilities,
for details see Hilgenfeldt \etal (1996).
Bubbles on the thick lines are in diffusive equilibrium
for $p_\infty/P_0=0.2,0.002$, respectively. Thick arrows indicate regions
of bubble growth
and shrinking by diffusive processes.
}
\label{total}
\end{figure}

Figure~\ref{total} illustrates how the conditions (i)--(iii) determine domain
boundaries
in the $P_a$--$R_0$ state space. Criterion (i) means that the Mach number
with respect to $c_{g}$ is larger than 1, i.e.,
\begineq
|M_{g}|= {|\dot R|\over c_{g} } \ageq 1,
\label{mach}
\endeq
and it is fulfilled for bubbles with large enough ambient radii $R_0$ and
at large enough forcing amplitude $P_a$, i.e., right of the dashed line in
figure~\ref{total}.
The shape stability condition (ii) -- see \cite{ple54,bir54,ell70,str71,pro77},
Brenner \etal (1995) or Hilgenfeldt \etal (1996) for detailed 
studies~\hbox{--,}
on the other hand,
limits the parameter domain in which bubbles can stably
oscillate to small $R_0\aleq 4-5\,\mu m$, within our boundary layer
approximation.
As the RP SL approach neglects
effects of thermal conduction, which has a damping influence on surface
oscillations, this upper limit on $R_0$ may be somewhat higher in reality.
Holt \& Gaitan's (1996) experimental results seem to give a threshold
around $7\,\mu m$. 
(i) and (ii) together determine the shaded
area of potentially sonoluminescing bubbles in figure~\ref{total}.
The actual position of a stable SL bubble in $P_a$--$R_0$
parameter space
is determined by condition (iii) for stable diffusive equilibria of the
gas inside the bubble and the dissolved gas
in the liquid (thick lines in figure~\ref{total}). These equilibrium lines
show negative slope whenever the equilibria are unstable, i.e., bubbles
below the line shrink and dissolve, bubbles above the line grow. At large
gas concentrations in the liquid (e.g.\ $p_\infty/P_0\sim 0.2$, left curve),
only unstable equilibria are possible in the parameter range of interest.
Tiny ratios $p_\infty/P_0\sim 0.002$ (right curve) are necessary
for diffusive stability (i.e., the fluid must be strongly degassed). The
positive slope of the upper branch of the curve characterizes these
bubbles as stable. The computation of diffusive
equilibria is explained in \S\,\ref{subsecdiff}.

\subsection{Summary of results of the present work}\label{secsummary}

Having identified the parameter regions for SBSL through numerically solving
the RP equation,
the question arises if one can understand the shape and size
of these regions
{\em analytically}, i.e., by analysing the bubble dynamics equations.
In principle,
all of the conditions that determine the occurrence of
stable/unstable/no SL depend only on properties of bubble dynamics.
Therefore, we set out in this work to derive analytical approximations
for RP dynamics and subsequently find scaling laws or approximate
analytical
expressions for our numerical curves presented above, in order to give a
clearer insight into
the role of different physical processes governing the dynamical equations.
Moreover, more practical reasons make analytical expressions highly desirable,
as the multi-dimensional parameter space of SBSL experiments cannot be
scanned in detail just by numerical solution of the RP equation. Our analytical
efforts strongly build on previous work, most notably that of
L\"ofstedt \etal (1993).
We present the most important results in this subsection, written such
that experimental parameters can be directly inserted to yield numerical
values. Here we have used fixed $\om=2\pi\times 26.5\,{\mbox{\it kHz}}$ and
$P_0=1\,atm$.
More detailed results and
the complete derivations for general $\om, P_0$
will be given in the corresponding Sections.
All the presented approximations naturally have limited parameter regimes
of validity, which include the region of sonoluminescing bubbles in all cases. 

We will demonstrate in \S\,\ref{secrp} that,
in order to understand the location of diffusive
equilibria, it is sufficient to analyse the parameter dependence of
the ratio of the maximum bubble radius to its ambient radius $(R_{max}/R_0)$,
see L\"ofstedt \etal (1993).
We show in \S\,\ref{sectwo} that two clearly distinct kinds of
bubble dynamical behaviour exist depending on $P_a$ and $R_0$:
{\em weakly oscillating} and {\em strongly collapsing} bubbles. The transition
between these two states is rather abrupt and occurs for given $P_a$ at
an ambient radius
\begin{equation}
R_0^{tr}={4\over 9}\sqrt{3}{\sigma\over P_a-P_0}
\approx {0.562\,\mu m \over P_a/P_0-1} \, .
\label{r0trintro}
\end{equation}
This transition is controlled by the surface tension $\sigma$, i.e., strong
collapses are easier to achieve for small $\sigma$.

In \S\,\ref{sectour}, we derive analytical approximations
to RP dynamics for all phases of the oscillation cycle of a strongly
collapsing bubble. We find that in this regime the bubble essentially
collapses like an empty cavity (see \cite{ray17}) according to
\begineq
R(t)\approx 14.3\,\mu m \cdot \left(R_{max}\over \mu m\right)^{3/5} 
\left({t^*-t\over T}\right)^{2/5}\, ,
\label{rayintro}
\endeq
with the time of maximum bubble compression $t^*$ and the driving period
$T=2\pi/\om$.

Following the collapse, a series of characteristic afterbounces of the
bubble radius occurs. We show in \S\,\ref{secafter} that they are the
cause for the wiggly structure of the diffusive equilibrium curves and the
$|M_{g}|=1$ line in figure~\ref{total}. The location of the wiggles can be
understood as a parametric resonance phenomenon. A Mathieu approximation
yields the ambient radius of the $k^{th}$ wiggle as
\begineq
R_0^{(k)}\approx 37.0\,\mu m \cdot
{q^{5/3}\over\sqrt{q^{5/3}-1}}{1\over k} +
0.487\,\mu m \cdot \left({q^{5/3}-2q+1\over q^{5/3}-1}+2{2-q^{2/3}\over
q^{2/3}}\right) \, 
\label{resintro}
\endeq
with the abbreviation $q=(1+P_a/P_0)$.

Section~\ref{secexp} deals with the bubble expansion. In the regime of strong
bubble collapses, an approximate result for the dependence of the
maximum radius on $P_a$ and $R_0$ is
\begineq
{R_{max}\over\mu m}\approx
67.2+0.112\left(R_0\over\mu m\right)^2+99.5(P_a/P_0-\pi/2)\, .
\label{practrmaxintro}
\endeq
With $R_{max}$, the location of diffusive equilibria in ($P_a,R_0$) parameter
space can be calculated.

A closer discussion of the role of surface tension and viscosity of the
liquid $\eta_l$ is
presented in \S\,\ref{secsurfvis}. In particular, the viscosity of
water is so small that it has no significant influence on bubble
dynamics. Oscillations are only viscosity-dominated if
\begineq
\eta_l^c\ageq \left(1+{0.487\,\mu m\over R_0}\right)\cdot 8.72 
\left({R_0\over\mu m}\right)\eta_{\em water} \, ,
\label{etacintro}
\endeq
which corresponds to $\eta_l\ageq40\eta_{\em water}$ for typical $R_0$.

Note that these equations are {\em not} fit formulas, but are analytically
{\em derived} from the RP dynamics. They are
all verified by comparison to full numerical solutions in the appropriate
domains of validity. 
With these formulas,
we are able to understand most of the parameter dependences of
SL analytically. Section \ref{secconcl} presents
conclusions.

\section{Rayleigh--Plesset bubble dynamics}\label{secrp}

\subsection{Notation and parameters}\label{subsecnot}
Since Lord Rayleigh (1917, see Lamb 1932 for earlier references)
treated the collapse of an empty
cavity in a liquid, a lot of refinement has been done in the modelling
of the dynamics of spherical domain walls in liquids. The main step
towards bubble dynamics was the introduction of
a variable external driving pressure and of the influence of surface tension
by \cite{ple49}.

An ODE for the bubble radius can be derived from the Navier--Stokes
equations from an approximation valid to the order of $\dot{R}/c_l$,
where $\dot{R}$ is the
speed of the bubble wall and $c_l$ is the sound speed in the liquid.
Following \cite{pro86}, L\"ofstedt \etal (1993) and many others, we will
henceforth denote the following ODE as Rayleigh--Plesset (RP) equation:
\begin{eqnarray}
\rho_l \left( R \ddot R + {3\over 2} \dot R^2 \right)  &=&
p_{gas}(R,t) - P(t) - P_0
             \nonumber \\
   &+& {R\over c_l} {{\mbox d}\over {\mbox d}t} p_{gas}(R,t) - 4 \eta_l 
{\dot R \over R} - {2\sigma \over R}.
		    \label{rp}
\end{eqnarray}
The left-hand side of this ODE for the bubble radius $R$ consists of dynamical
pressure terms already known to Rayleigh ($\rho_l= 1000\, kgm^{-3}$ is the
density of water). $P_0=1\,atm$ is the constant
ambient pressure, $P(t)$ the ultrasound driving, modelled as a
spatially homogeneous, standing
sound wave, i.e.,
\begineq
P(t) = - P_a \cos \om t = - P_0\cdot p \cos \om t
\label{eq1}
\endeq
with the dimensionless forcing pressure amplitude $p\equiv P_a/P_0$ and
a fixed frequency of $\om=2\pi\times 26.5\,{\mbox{\it kHz}}$
(period $T\approx 38\,\mu s$),
which is a common value in
many experiments like those of \cite{bar94} and Hiller \etal (1992).
The wavelength of this sound in water is
about $5\,cm$, while the bubble radii treated in this work never exceed
$200\,\mu m$. Because of this separation of scales, it is common to assume
spatial homogeneity, as stated above.
We will refer to the sum of experimentally controllable pressures
as the {\em external pressure} $p_{ext}=P_0+P(t)$. By definition,
the external pressure exerts maximally outward directed forces
($p_{ext}=P_0(1-p)<0$) on the bubble at $t=0$.

The other terms on the right-hand side of equation~(\ref{rp}) model the
influence of the surface tension at the bubble-water interface 
($\sigma = 0.073\,kg\,s^{-2}$), the water viscosity 
($\eta_l = 1.00\times 10^{-3}\,{\mbox{\it Pa}}\, s$), 
and of emitted sound waves from the bubble
(cf.\ Keller \& Miksis 1980,
this term contains the speed of sound in water $c_l=1481\,m\, s^{-1}$).

%caption2
\begin{figure}[htb]
\setlength{\unitlength}{1.0cm}
\begin{picture}(12,15.2)
\put(-0.,5.4){\psfig{figure=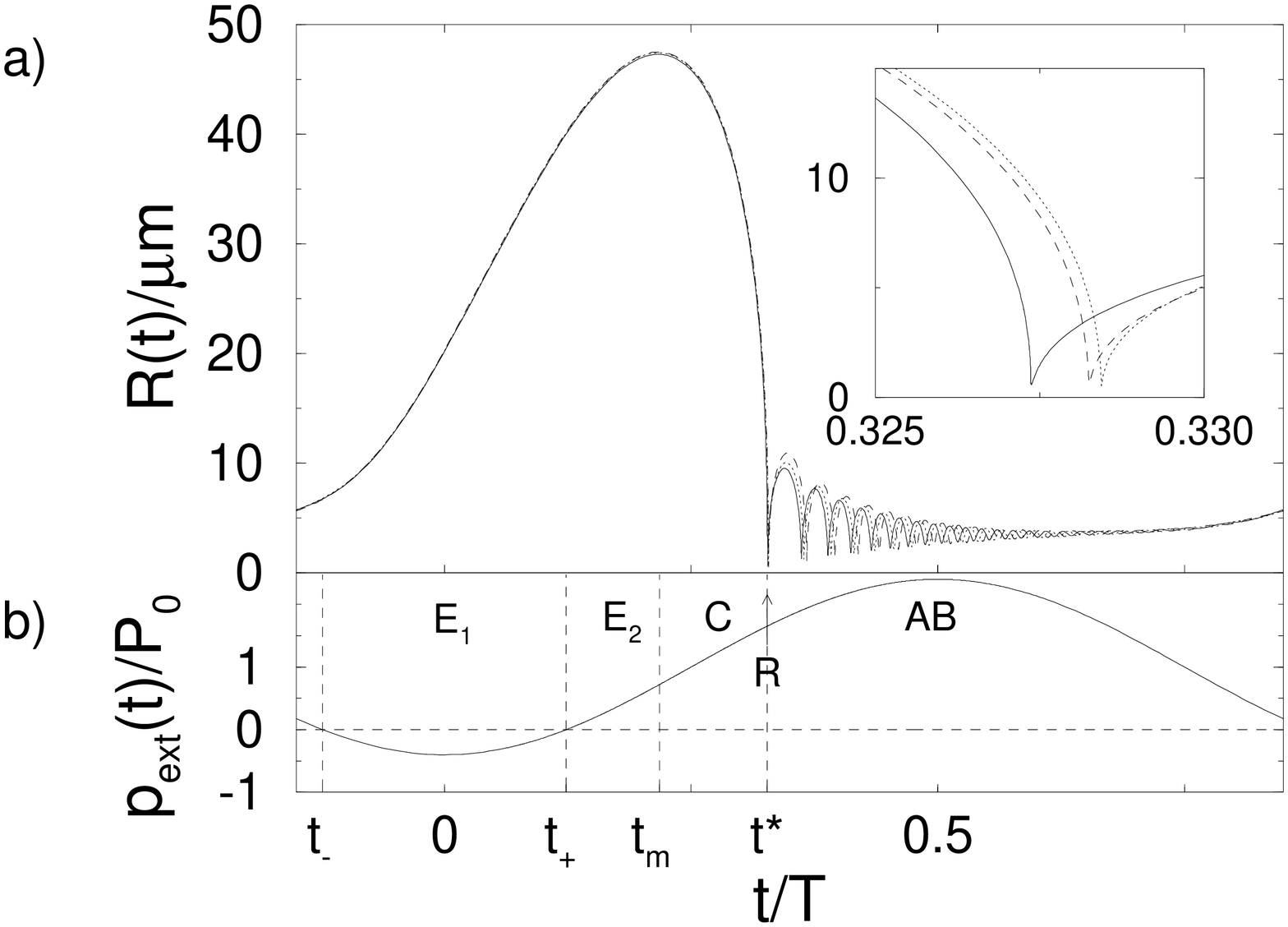,width=12cm,angle=-90.}}
\put(0,-4.1){\psfig{figure=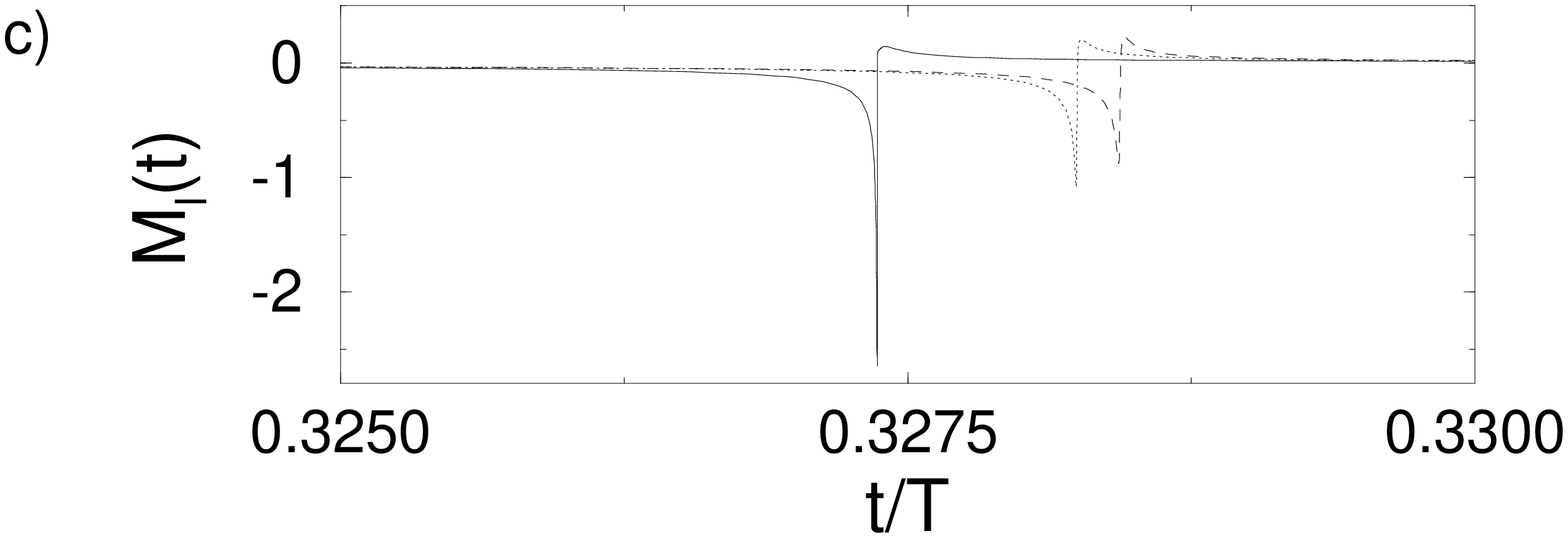,width=12cm,angle=-90.}}
\end{picture}
\caption[]{($a$) Bubble dynamics for $P_a=1.4\,atm$, $R_0=4.0\,\mu m$ resulting
from
the RP equation (\ref{rp}) (solid), Flynn's equation (\ref{flynn}) (dotted)
and Gilmore's equation (\ref{gilmore}) (dashed). The inset shows
a blowup of the vicinity of the collapse.
($b$) External pressure $p_{ext}$ for the dynamics in ($a$).
$t_m$ is the time of
maximum expansion, $t^*$ the time of collapse; $p_{ext}=0$ at $t=t_+,t_-$.
The different intervals of the oscillation cycle treated in Section
\ref{sectour} are indicated. ($c$) Mach numbers $M_l$ for the time interval
displayed in the inset of ($a$).
For the RP equation (solid) and Flynn's equation (dotted)
$M_l=\dot{R}/c_l$ with constant $c_l$, for 
Gilmore's equation (dashed) $M_l=\dot{R}/C_l$ with pressure dependent $C_l$.
}
\label{roft}
\end{figure}

The gas pressure $p_{gas}(R,t)$ inside the bubble is assumed to obey a
van der Waals type process equation
\begineq
p_{gas}(R,t) = p_{gas}(R(t)) = \left(P_0 + {2\sigma \over R_0}\right)
\left( {R_0^3 - h^3\over R^3(t) - h^3 }\right)^\kappa,
\label{vdw}
\endeq
$R_0$ being the ambient bubble radius and $h$ the (collective) van der Waals
hard core radius $h= R_0/8.86$ (for argon) (Lide 1991). The pressure exerted
by surface tension was included explicity in (\ref{rp}). The 
$\sigma$ dependence of the prefactor of
the polytropic expression ensures that $R_0$ is the radius of a
static (unforced) bubble, neglecting effects of gas diffusion.
Note that (\ref{vdw}) presupposes
homogeneity of the pressure inside the bubble. This is of course not satisfied
in the final stages of bubble collapse,
as a more detailed investigation of the gas dynamics inside the bubble reveals,
cf.\ \cite{wu93,mos94,vuo96,eva96,bre96d}, Moss \etal (1997),
but the violent collapse phase lasts only $\sim 1\,ns$ out of the
$T\approx 38\,\mu s$ of the oscillation cycle. Therefore, this approximation
does not severely affect our analysis of bubble wall
dynamics. We furthermore set the effective polytropic 
exponent $\kappa \approx 1$ as for this frequency and bubble ambient
radii below $\sim20\,\mu m$ the bubbles can be considered to be isothermally
coupled to the surrounding liquid (Plesset \& Prosperetti 1977),
except during the small time
interval around the bubble collapse, where the
extremely rapid bubble dynamics requires adiabatic treatment of the gas.
This will be taken into account in \S$\!$\S\,\ref{seccollapse} and \ref{turn}.
The solid line of figure~\ref{roft}($a$) shows a time series $R(t)$ from
(\ref{rp})
for relatively strong driving $P_a=1.4\,atm$ and moderate ambient radius
$R_0=4.0\,\mu m$. 
The typical feature of the oscillations of $R(t)$ is a slow expansion for
approximately half a cycle of driving,
followed by a rapid and violent collapse and a series of afterbounces
corresponding to an almost free oscillation of the bubble.
The time scale of the afterbounces is thus set by the period of the bubble's
(small amplitude) eigenoscillations, whose frequency $\om_e\sim 1\,{\mbox
{\it MHz}}$
can be easily
obtained from a linearization of (\ref{rp}):
\begineq
\ddot{R}+\om_e^2 (R-R_0)={P_a \cos \om t\over\rho_l R_0} \quad {\mbox{with}}
\label{fulllin}
\endeq
\begineq
\om_e^2={3P_0\over \rho_l R_0^2} \, ,
\label{ome}
\endeq
where we have set $\kappa=1$ and neglected
surface tension and viscosity effects.
Including surface tension yields
\begin{equation} 
\om_s^2={3P_0\over\rho_l R_0^2}+{4\sigma\over \rho_l R_0^3} =\left(1+{2\over
3}\alpha_s\right) \om_e^2  \, ,
\label{omes}
\end{equation}
where $\alpha_s=2\sigma/(P_0 R_0)$ is the ratio of surface tension pressure
to $P_0$ at
$R=R_0$. $\alpha_s\approx1$ for
$R_0\approx 1.5\mu m$, while for larger $R_0$ it becomes very small.

The {\em resonance radius}, on the other hand, is defined as the
ambient radius of a bubble with $\om_e=\om$, i.e., 
\begineq
R_{res}=\left({3P_0\over\rho_l \om^2}\right)^{1/2}\approx 105\mu m \, .
\label{rres}
\endeq

\begin{table}
  \begin{center}
 \begin{tabular}{ll}
%  \hline
  {pressure term} & {definition} \\[3pt]
%%  \hline
 $p_{acc}$ & $\rho_l R\ddot{R}$ \\
 $p_{vel}$ & ${3\over 2}\rho_l\dot{R}^2$ \\
 $p_{gas}$ & $\left(P_0+{2\sigma\over R_0}\right)\left({R_0^3-h^3\over R^3-h^3}
 \right)^\kappa$ \\
 $p_{sur}$ & ${2\sigma\over R}$ \\
 $p_{vis}$ & $4\eta_l{\dot{R}\over R}$\\
 $p_{snd}$ & ${R\over c_l} \left(P_0+{2\sigma\over R_0}\right)
 {{\mbox d}\over {\mbox d}t}
  \left({R_0^3-h^3\over R^3-h^3}\right)^\kappa$ \\
 $p_{ext}$ & $P_0-P_a\cos\om t = P_0(1-p\cos\om t)$
%%  \hline
 \end{tabular}
  \end{center}
\caption{Definition of the pressure terms in the
RP equation (\protect\ref{rp}) used in this work.}
\label{table1}
\end{table}

For convenience, we list in table \ref{table1}
the definition of the different pressure
terms of (\ref{rp}) which will appear throughout this paper.

Besides the solution of the RP equation (\ref{rp}), figure~\ref{roft}($a$)
shows time series obtained from other commonly used bubble dynamical
equations, namely Flynn's and Gilmore's equation, which are discussed in
detail in Appendix A. It is obvious that, for bubbles in the SBSL regime,
all equations yield very similar $R(t)$ dynamics. It is only upon
magnification of the small time interval around the collapse
(figure~\ref{roft}$b$) that the differences between these descriptions of
bubble dynamics becomes apparent.

The deviations of the RP, Flynn, and Gilmore equations from each other
may become pronounced when the bubble is driven
at very high pressure amplitudes such as $P_a=5\,atm$ (cf.\ Lastman \& Wentzell
1981).
These pressures are common in cavitation fields, but they are far too high
to allow for stable bubbles in SBSL experiments (with the possible exception
of SBSL in high magnetic fields described by Young, Schmiedel \& Kang 1996).
Sonoluminescent bubbles
require a driving pressure amplitude in a narrow window
$1.1\,atm$$\aleq P_a\aleq 1.5\,atm$. It is this range of $P_a$ that we will 
mainly focus on in this work.
Only in \S\,\ref{secexp}
results in the range of cavitation field pressures will briefly be displayed.
Direct and indirect measurements of the
size of SL bubbles e.g.\ in \cite{bar92}, Tian \etal (1996), or \cite{hol96}
indicate that typical $R_0$ lie
around $5\,\mu m$.

\subsection{Calculating diffusive equilibria from RP dynamics}
\label{subsecdiff}
A computation of points of diffusive equilibrium in the $P_a$--$R_0$ plane
from first principles requires
solution of an advection diffusion PDE with appropriate boundary conditions,
coupled to the RP equation. This is numerically far too expensive to allow
for a scan of the whole $P_a$--$R_0$ parameter space.
In \cite{bre96} and Hilgenfeldt \etal (1996),
we therefore employed the method introduced by \cite{fyr94} and
\cite{loe95},
which is based on the separation of the driving time scale
$T$ and the diffusive time scale $\tau_{\em diff}\gg T$.
Within this approximation, the task is massively
reduced to the solution of the RP equation and the computation of weighted
averages of the form
\begineq
\left< f \right>_i = \int_0^T f(t) R^i(t) {\mbox d}t \left/ 
\int_0^T R^i(t) {\mbox d}t \right. .
\label{av}
\endeq
The mass flux into or out of the bubble is then proportional to
$p_\infty - \left< p_{gas} \right>_{4}$ (see Fyrillas \& Szeri 1994).
An equilibrium
point is characterized by the simple condition    
\begineq
p_\infty = \left< p_{gas} \right>_{4}
\label{diffeq}
\endeq
and it is stable if
\begineq
\beta =  {d\left< p_{gas} \right>_{4}\over dR_0}
\label{diffstab}
\endeq
is positive.

%caption3
\begin{figure}[htb]
\setlength{\unitlength}{1.0cm}
\begin{picture}(12,11)
\put(-0.5,0.5){\psfig{figure=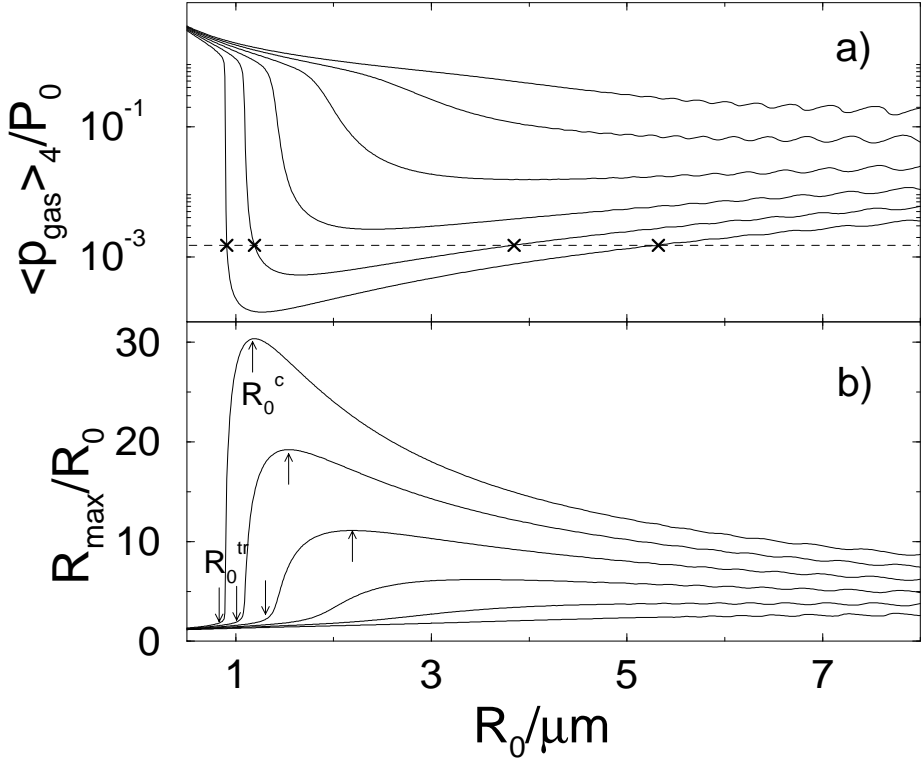,width=13cm,angle=-90.}}
\end{picture}
\caption[]{($a$) Average pressure $\la p_{gas} \ra_{4}$.
The curves are parametrized with the dimensionless driving pressures
$p=1.0,\,1.1,\,1.2,\,1.3,\,1.4,$ and 1.5 from top to bottom. Note the
logarithmic
scale of the ordinate. The horizontal dashed line is an example for a given
gas concentration $p_\infty/P_0$ in the liquid. Crosses mark the corresponding
$R_0$ where diffusive equilibria are found.
($b$) Expansion ratio $R_{max}/R_0$ for the same
$p$ as in ($a$): $p=1.0$ to $1.5$ in steps of 0.1 from bottom to top. 
The maxima occur at almost exactly the same $R_0=R_0^c$ as the minima of
$\left< p_{gas}\right>_{4}$. Up arrows mark the $R_0^c$,
down arrows indicate the transition radii $R_0^{tr}$ for p=1.3,\,1.4,\,1.5.
}
\label{p4}
\end{figure}

Figure~\ref{p4}$a$ displays $\left< p_{gas} \right>_{4}$ for different $P_a$.
The graphs show characteristic wiggles for larger $R_0$ (which can be
explained
from resonance
effects, see \S\,\ref{secafter}) and, for large enough $P_a$, a global
{\em minimum} at some critical $R_0=R_0^c$. If $R_0>R_0^c$, even with no
wiggles present, the bubbles are diffusively stable according to the sign of
the slope $\beta$.
For small $R_0<R_0^c$, all equilibria are unstable, i.e., the bubble either
dissolves or grows by rectified diffusion, see \cite{bla49,ell64}
(the latter case can lead to
unstable SBSL, cf.\ Hilgenfeldt \etal 1996).
The possibility of multiple stable equilibria because
of the resonance structure was recognized earlier by \cite{chu88} and
\cite{kam93}.
Here we analyse the formal and physical origin of the positive
{\em overall} slope of
$\left< p_{gas}\right>_{4}(R_0)$ for large $R_0$, which is an essential
property of stable SBSL bubbles.

In the average $\left< p_{gas} \right>_{4}$ the pressure is weighted with
$R^4(t)$ and will therefore
be dominated by the value of $p_{gas}$ at $R_{max}$. For large radii, we can 
neglect the excluded volume $h^3$ in the van der Waals formula and replace
(\ref{vdw}) 
by an ideal gas law under isothermal conditions,
\begineq
{\left< p_{gas} \right>_{4}\over P_0} \approx
\left(1+\alpha_s\right)
{\int_0^T R_0^3R(t) {\mbox d}t \over \int_0^T R^4(t) {\mbox d}t }\approx
\xi\left(1+\alpha_s
\right) \left({R_0\over R_{max}}\right)^3 \, .
\label{diffp4rmax}
\endeq
$\xi$ is a prefactor that is due to the different shape of the
integrands $R(t)$ and $R^4(t)$. A crude estimate of $\xi$ can be obtained
by approximating $R(t)$ by a parabola $\widetilde{R}(t)\sim
R_{max}(1-16t^2/T^2)$ and integrating
$\widetilde{R}$ and $\widetilde{R}^4$ over one half cycle from $-T/4$
to $T/4$. This gives
$\xi=105/64\approx 1.64$, which is quite accurate in reproducing numerical
results.
With this saddle point approximation introduced by L\"ofstedt \etal (1993),
the key parameter for diffusive
equilibria is the expansion ratio $R_{max}/R_0$. Figure~\ref{p4} demonstrates
the close relation between $\left< p_{gas} \right>_{4}$ and
$R_{max}/R_0$ as functions of $R_0$.
The expansion ratio displays a maximum at $R_0^c$,
corresponding to the minimum of
$\left< p_{gas}\right>_{4}$. In order to determine diffusive equilibrium
points, one has to look for the intersections of the
$\left< p_{gas}\right>_{4}/P_0$ curves in figure~\ref{p4} with a horizontal
line given by $p_\infty/P_0$ (cf.\ equation~(\ref{diffeq})). Note that degassing
to tiny partial pressures is necessary to achieve equilibria in the
$R_0$ range of pure argon SL bubbles; this fact was first realized by
\cite{loe95}.

For high enough $P_a$,
there are two equilibrium values for $R_0$, the larger one being a stable
equilibrium, the smaller one being unstable.
If $P_a$ is decreased, $\left< p_{gas}\right>_{4}/P_0$ increases and
the equilibria come closer together. This can also be seen in
figure~\ref{total}:
for decreasing $P_a$, the $R_0$ values given by the $p_\infty/P_0=0.002$
equilibrium curve approach each other. Eventually, at a certain $P_a$ the
stable und the unstable equilibrium coalesce and for smaller $P_a$ no
equilibrium is possible. This is reflected in figure~\ref{p4} by the fact
that the whole $\left< p_{gas}\right>_{4}/P_0$ curve lies above
$p_\infty/P_0$.

For relatively high gas concentrations such as $p_{\infty}/P_0=0.2$,
stable equilibria can only exist for very large $R_0$, where the
bubbles are shape unstable. But if the concentration is lowered, e.g.\
to $p_{\infty}/P_0=0.002$, the stable branch (positive slope in
figure~\ref{total}) enters the region of
sonoluminescent bubbles, whereupon stable SL can set in.
The occurrence of stable and unstable branches depends on the existence of
a minimum in $\left< p_{gas}\right>_{4}$, which in turn necessitates
a maximum in $R_{max}/R_0$ (figure~\ref{p4}$a$ and $b$).
Therefore, to analyse the lines of diffusive equilibria in figure~\ref{total},
it is sufficient to explain the maximum of the expansion 
ratio figure~\ref{p4}($b$)
and its dependence on $R_0$ and $p$; this question will be addressed in
\S\,\ref{secexp}.

\section{Quasistatic Blake threshold}\label{sectwo}

\begin{figure}[htb]
\setlength{\unitlength}{1.0cm}
\begin{picture}(12,11)
\put(-0.5,0.5){\psfig{figure=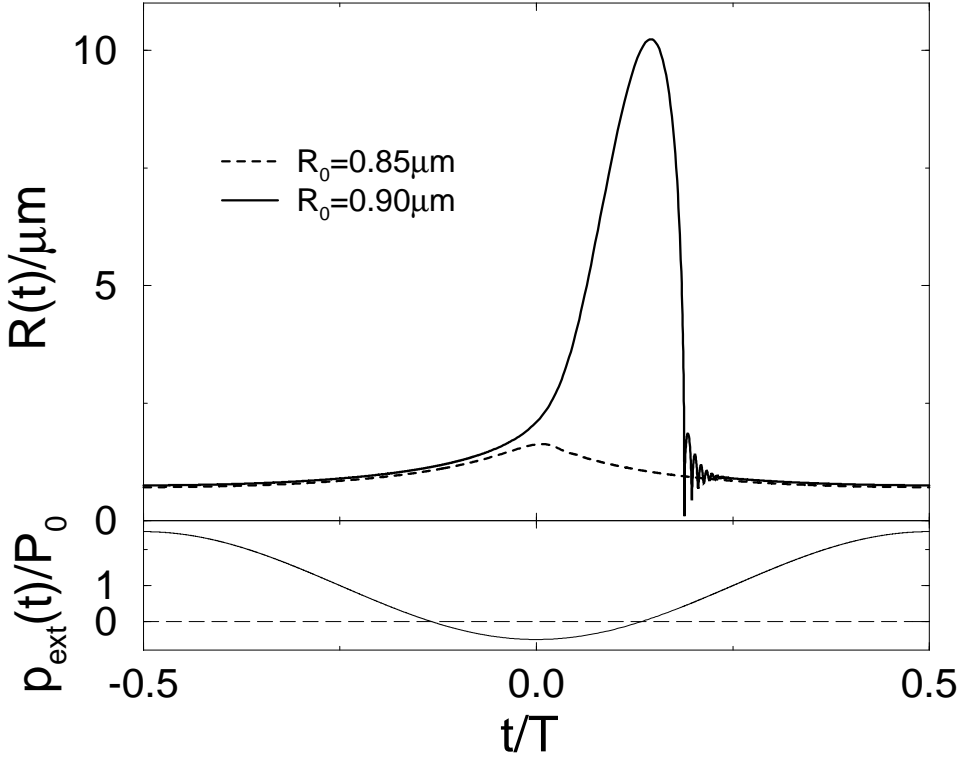,width=13cm,angle=-90.}}
\end{picture}
\caption[]{Bubble dynamics for $P_a=1.5\,atm$, $R_0=0.85\,\mu m$ (dashed) and
$R_0=0.90\,\mu m$ (solid). The larger bubble undergoes dynamical expansion
and a strong collapse, the smaller oscillates weakly with the driving field
(lower part). 
}
\label{strongweak}
\end{figure}

The transition from sharply increasing $R_{max}/R_0$ for small $R_0$ to
decreasing expansion ratios for large $R_0$ (figure~\ref{p4}$b$) marks an
important boundary between two very different types of bubble dynamics.
Consider figure~\ref{strongweak} where two 
examples of bubble dynamics for the same $P_a=1.5\,atm$
and only minutely different ambient radii are displayed. The smaller
bubble exhibits a weak (although obviously not sinusoidal) oscillation with
a maximum expansion ratio $R_{max}/R_0\approx 2$;
no collapse is visible. The time series of the larger bubble is almost
indistinguishable from the other until $t\approx 0$. But then,
a rapid expansion to $R_{max}/R_0\approx 10$ occurs, followed by a strong
collapse, the typical dynamics of a sonoluminescing bubble, cf.\
figure~\ref{roft}($a$).

%caption4

%caption5
\begin{figure}[htb]
\setlength{\unitlength}{1.0cm}
\begin{picture}(12,8)
\put(-6.2,-11.2){\psfig{figure=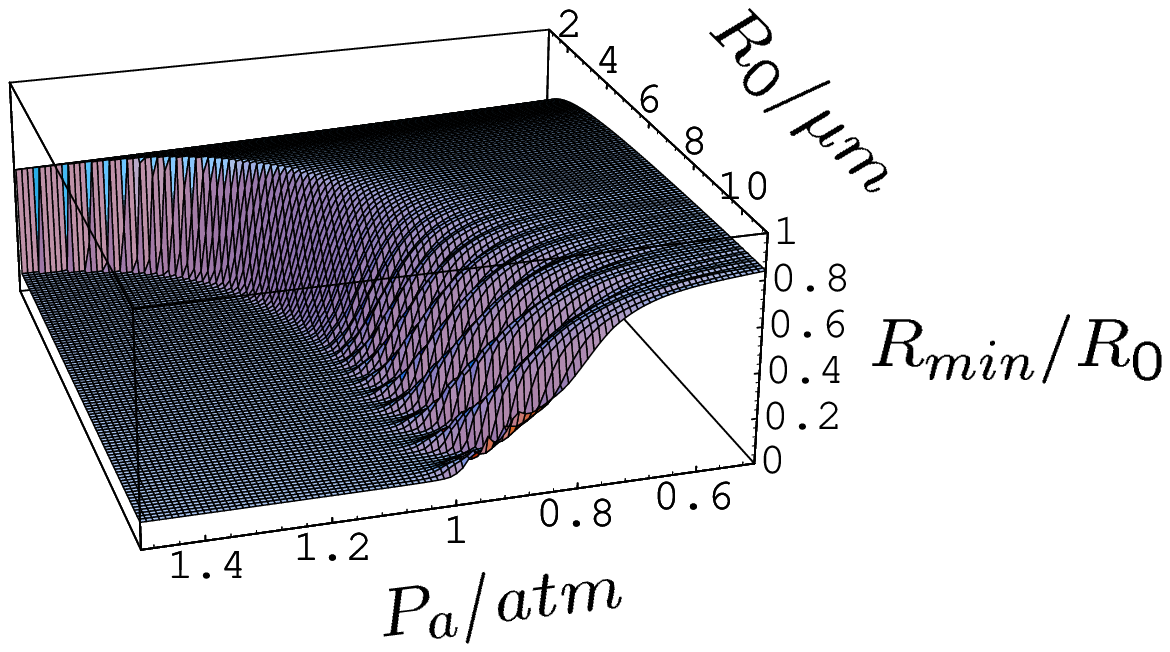,width=20cm,angle=0.}}
\end{picture}
\caption[]{Compression ratio $R_{min}/R_0$ as a function of $P_a$ and
$R_0$. The two regimes of bubble dynamics are clearly visible: weakly
oscillating bubbles for small $R_0$ and small $P_a$, strong collapses
to the hard core radius for large $R_0$ and large $P_a$.  
}
\label{rminr0}
\end{figure}

Figure \ref{rminr0} shows the compression ratio $R_{min}/R_0$ of the minimum
radius achieved during bubble oscillation to the ambient radius
as a function of $P_a$ and $R_0$. A sharp transition, like in the expansion
ratio, is obvious in this graph and it occurs at the same $R_0$.
For small $P_a$ and small $R_0$, $R_{min}/R_0$ is near one; we denote such
bubbles as {\em weakly oscillating}. For large $P_a$ and $R_0$, a horizontal
plane at $R_{min}/R_0\approx h/R_0$ indicates collapse to a radius very 
near the hard core radius. We say that these latter bubbles exhibit
{\em strong collapses}.

The key to understanding this transition from weakly oscillating
to strongly collapsing bubbles lies in the existence of a threshold for
spontaneous bubble expansion known as the Blake threshold
(Blake 1949, Atchley 1989).
It is normally considered for bubbles under static conditions: let us
first set $P_a$ (and thus also $p_{ext}$) constant in time, and
correspondingly take $R(t)$ to be time-independent. Then the RP equation
reduces to
\begin{equation}
0 = (P_0 + {2\sigma \over R_0})
\left( {R_0\over R}\right)^3
 - p_{ext} - {2\sigma \over R}\, ,
		    \label{blake}
\end{equation}
where for $p_{gas}$ again the isothermal ideal gas law was used,
which is certainly an excellent
approximation for the static situation.
For $p_{ext}>0$, equation~(\ref{blake}) has exactly one solution for
positive $R$,
and it corresponds to a stable equilibrium. If $p_{ext}<0$ but small in
absolute magnitude, two equilibria exist, the one at larger $R$ being
unstable, i.e., a bubble with larger radius would grow indefinitely.
Finally, at a critical $p_{ext}^B<0$ (Blake threshold pressure, cf.\
Prosperetti 1984) the two
equilibrium points merge and
disappear in an inverse tangent bifurcation. In this situation, $p_{gas}$
is always larger than $p_{ext}+p_{sur}$ and (\ref{blake}) cannot
be fulfilled for any radius. Thus, the assumption of a time-independent
$R(t)$ has to be dropped.
A dynamical expansion ensues with significant contributions
from the dynamical pressure terms on the left-hand side of (\ref{rp}).

Returning to the oscillatory driving $p_{ext}=P_0(1-p\cos\om t)$, we notice
that the driving period $T=2\pi/\om\approx 40\,\mu s$ is long compared to
the time scale of the bubble's eigenoscillations
$2\pi/\om_e\sim 1\,\mu s$. 
Thus, we
can consider the external pressure oscillations as quasistatic and follow
Blake's argument as above. As $p_{ext}<0$ is necessary to cross the Blake
threshold, we must require $p>1$ here.
Obviously, the most sensitive point in the cycle
is $t=0$, where $p_{ext}$ is negative and of magnitude $(p-1)P_0$.

\begin{figure}[htb]
\setlength{\unitlength}{1.0cm}
\begin{picture}(12,11)
\put(0.5,0.5){\psfig{figure=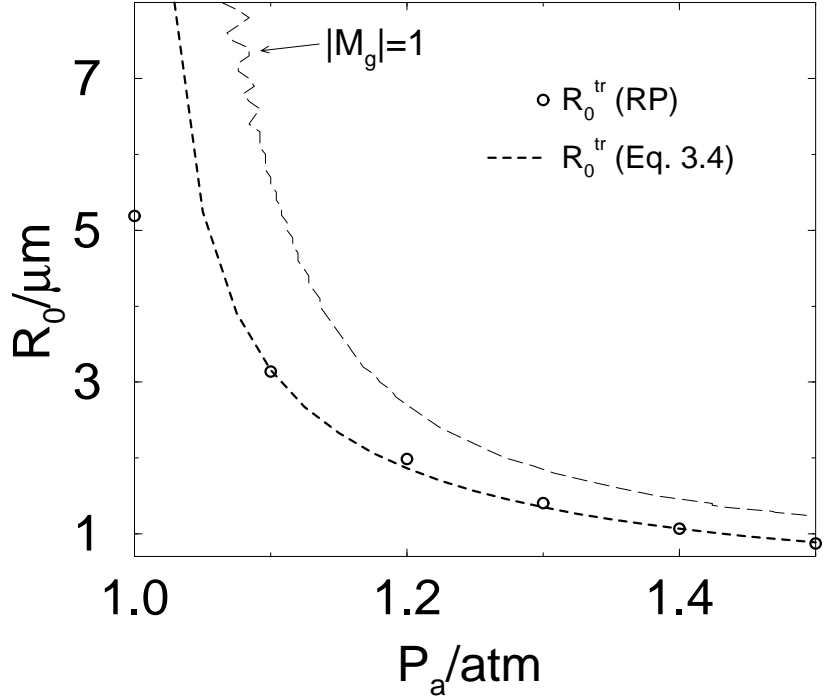,width=13cm,angle=-90.}}
\end{picture}
\caption[]{Transition ambient radii $R_0^{tr}$ from numerical solution of
the RP ODE (circles) and from 
(\ref{r0trans}) (dashed).
This figure shows the same parameter range as figure~\ref{total},
from which the $|M_{g}|=1$ curve was taken (thin dashed line).
The transition to collapsing bubbles occurs for slightly smaller pressures and
ambient radii than the onset of light emission at $|M_{g}|=1$.
}
\label{transition}
\end{figure}

The quasistatic approximation (\ref{blake}) describes the {\em complete}
time series of a weakly oscillating bubble with good accuracy.
Rewriting (\ref{blake}), we obtain the cubic equation
\begineq
(p\cos\om t-1)R^3-{2\sigma\over P_0}R^2+
\left(1+{2\sigma\over R_0P_0}\right)R_0^3 = 0 \, .
\label{cubicr}
\endeq
Given a time $t$ for which $p_{ext}<0$, there is a critical
$R_0=R_0^{tr}$ above which the two positive real 
solutions of (\ref{cubicr}) become complex. When this happens, the
weak oscillation dynamics is no longer a valid description and the transition
to strong collapses occurs.
For given $p$, the smallest transition radius $R_0^{tr}$ is
required for $t=0$. For $R_0^{tr}$, therefore, the discriminant of
(\ref{cubicr}) at $t=0$ must vanish, i.e.,
\begineq
R_0^3+{2\sigma\over P_0}R_0^2-{32\over 27}{\sigma^3\over P_0^3(p-1)^2} = 0.
\label{discriminant}
\endeq
After a lengthy but straightforward calculation, the transition ambient radius
$R_0^{tr}$ at given $p=P_a/P_0$ is
\begin{eqnarray}
R_0^{tr} &=& {2\over 3}{\sigma\over P_0}\left\{
\left(
{2\over (p-1)^2}-1+{2\over (p-1)}\sqrt{{1\over (p-1)^2}-1} \;\,
\right)^{1/3} \right. \nonumber \\
& + & \left.\left(
{2\over (p-1)^2}-1+{2\over (p-1)}\sqrt{{1\over (p-1)^2}-1}\;\,
\right)^{-1/3}
-1\right\} .
\label{r0trans}
\end{eqnarray}
Note that $R_0^{tr}$ is a real number for all $p$.
In figure~\ref{transition} the calculated $R_0^{tr}$ from (\ref{r0trans})
is compared to the numerical values (identified
by the condition $R_{min}/R_0=0.5$). The agreement is very good, the errors
at higher $P_a$ being only about $0.01\,\mu m$.

When $R_0$ exceeds $R_0^{tr}$, there is a period of time around $t=0$
where the right-hand side of (\ref{blake}) cannot be zero, but must be 
positive.
Then, the dynamical terms neglected so far must become noticeable and
a dynamical expansion follows which can only be stopped when $p_{ext}$
has again become large enough to allow for a stable radius equilibrium.
When the bubble growth is stopped,
the expanded bubble does not experience significant outward directed forces
and, consequently, undergoes a violent collapse. If $R_0$ is only slightly
larger than $R_0^{tr}$, the time scale separation still holds for a large
portion of the cycle, cf.\ figure~\ref{strongweak}.

%caption6

It is immediately obvious from (\ref{discriminant}) and (\ref{r0trans}) that
surface tension plays a key role in this transition mechanism from weak
oscillations to strong collapses.
If $p>1$, weak oscillations at small
$R_0$ are dominated by the
influence of $\sigma$, whereas strongly collapsing (larger) bubbles
are controlled by the properties of 
dynamical expansion and collapse
(cf.\ \S$\!$\S\,\ref{seccollapse},\,\ref{secexp}). Note that in a fluid with
very small $\sigma$, already bubbles of very small size will show collapses
(see also L\"ofstedt \etal 1995 and Akhatov \etal 1997).
It should also be emphasized here that the crucial driving parameter for
the transition is $(p-1)$, i.e., the {\em difference} of driving pressure
amplitude $P_a$ and ambient pressure $P_0$, rather than $P_a$ itself.

\begin{figure}[htb]
\setlength{\unitlength}{1.0cm}
\begin{picture}(12,11)
\put(0.,0.5){\psfig{figure=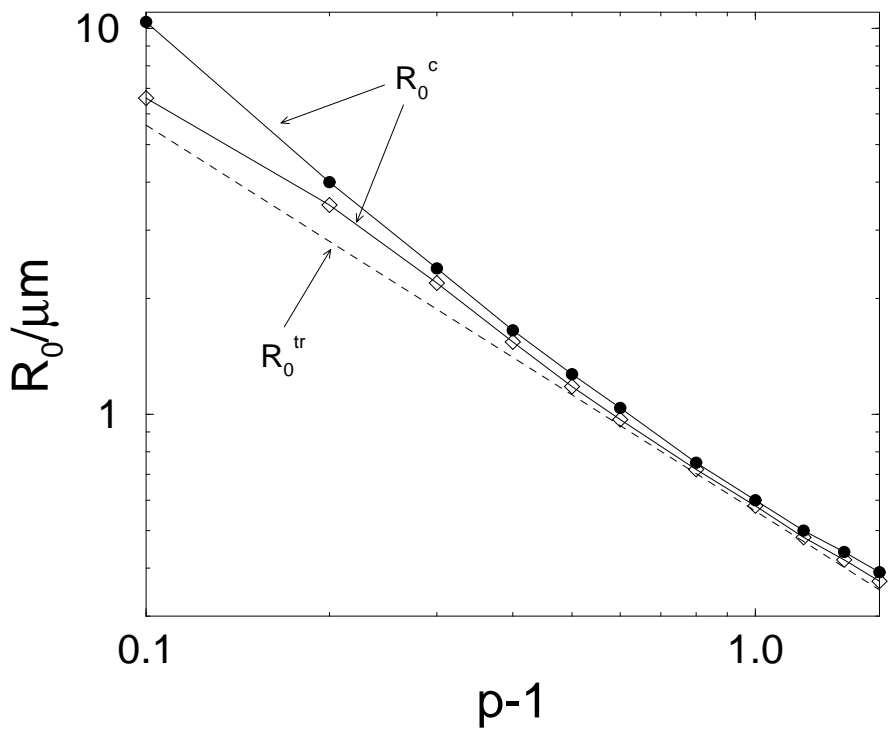,width=13cm,angle=-90.}}
\end{picture}
\caption[]{Critical ambient radii $R_0^c$
from numerical computation of the extrema of $\left< p_{gas}\right>_{4}$
(filled circles) and
$R_{max}/R_0$ (open diamonds) and from the asymptotic law for
$R_0^{tr}$ equation (\ref{r0tlargep}) (dashed line).
The scaling behaviour $R_0^c\propto 1/(p-1)$ from (\ref{r0tlargep})
is quite accurate for higher pressures $p\ageq 1.3$. 
}
\label{r0c}
\end{figure}

In the limit of large forcing $p\gg 1$,
(\ref{discriminant}) yields the much simpler formula
\begin{equation}
R_0^{tr}={4\over 9}\sqrt{3}{\sigma\over P_0}{1\over p-1} \, .
\label{r0tlargep}
\end{equation}
It can be seen from figure~\ref{p4} that in this limit the difference between
$R_0^{tr}$ (onset of transition) and $R_0^c$ (extremum of expansion and
compression ratio) becomes negligibly small. Thus, (\ref{r0tlargep}) is
also a good approximation to the critical $R_0^c$ we were trying to
identify. This is confirmed by figure~\ref{r0c}, from which also the (small)
errors of the saddle point approximation
(determining $R_0^c$ from $R_{max}/R_0$ instead of $\la p_{gas}\ra_{4}$)
can be read off.

%caption7

What is the maximum radius of a bubble weakly oscillating at $R_0^{tr}(p)$?
Inserting (\ref{r0tlargep}) into (\ref{cubicr}) with $t=0$ and
expanding to the same order in $1/(p-1)$ gives
\begin{equation}
R_{max}={4\over 3}{\sigma\over P_0}{1\over p-1}
\label{rmaxt}
\end{equation}
in the large $p$ limit.
This yields a minimum expansion ratio of $R_{max}/R_0^{tr}=\sqrt{3}$ for
the onset of bubble collapse, which is an analytical
justification of Flynn's (1975b) definition of a {\em transient cavity}.
In that work, a strongly collapsing bubble was characterized by an expansion
ratio $\ageq 2$. 

As the collapse sets in rather abruptly when $R_0$ is enlarged, we expect
that $R_0^{tr}$ also marks the transition to bubbles which fulfill the
Mach criterion (\ref{mach}). Figure~\ref{transition} shows  the
$|M_{g}|=1$ line of figure~\ref{total} together with the $R_0^{tr}(p)$ line
according to (\ref{r0trans}). Both curves display the same trend,
approaching each other at large $p$.
The Blake transition occurs for smaller $P_a$ and $R_0$ than those
necessary for $|M_{g}|\ageq 1$, i.e., for possible light emission.
The physical consequence of this is that, upon increasing the driving force,
the bubble first emits cavitation noise due to collapses and only afterwards
starts to emit light. Indeed, such a sequence of events has been reported
by W.~Eisenmenger \& B.~Gompf (private communication, 1996).

The transition line $R_0^{tr}(p)$ is shifted towards smaller $R_0$ for smaller
$\sigma$. This means that collapses of the same violence can be achieved
(for a given $R_0$ range) with smaller driving pressures in a liquid with less
surface tension.
Note however that such bubbles will also be stronger affected by surface
instabilities, whereas in a liquid with high $\sigma$, bubbles are more
surface stable. It is therefore possible to obtain violent collapses
at larger $R_0$ in liquids with larger surface tension
using larger driving pressures.

\section{A guided tour of RP dynamics}\label{sectour}

Let us now explain in detail the dynamics of strongly collapsing bubbles
(as shown e.g.\
in figure~\ref{roft}$a$). To this end, we divide the oscillation cycle
of the bubble into several time intervals indicated in
figure~\ref{roft}($b$), where
$t_m$ is the time of maximum bubble radius, $t^*$ the time
of minimum bubble radius (after collapse), and $t_+=-t_-=\arccos(1/p)/\omega$
the time when
$p_{ext}$ changes its sign from positive (contracting) to negative
(expanding) values. With this interval division scheme we extend an approach
presented in the pioneering paper by L\"ofstedt \etal (1993).
In particular, we will treat the {\em bubble collapse phase} denoted by C in
figure~\ref{roft}($b$) in the interval $t_m\leq t\leq t^*$, the
{\em reexpansion interval} (R) very close to the time of maximum compression 
($t\approx t^*$), the {\em afterbounces} (AB) for $t^*\leq t\leq t_-$ and
the {\em bubble expansion} in two stages for $t_-\leq t \leq t_+$ (E$_1$)
and $t_+\leq t \leq t_m$ (E$_2$).

Within each of these intervals, certain pressure terms in (\ref{rp}) are
dominant, whereas others are negligible. Thus, simplified equations
with analytical solutions can be derived, which enable us to characterize
the complex bubble behaviour analytically and quantitatively. Our approximate
formulas hold in the regime of strongly collapsing bubbles, i.e., for
$R_0>R_0^{tr}(P_a)$; in the weakly oscillating regime, the bubble dynamics 
becomes of course trivial.

\subsection{Rayleigh collapse (region C)}\label{seccollapse}

\begin{figure}[htb]
\setlength{\unitlength}{1.0cm}
\begin{picture}(12,11)
\put(0.,0.5){\psfig{figure=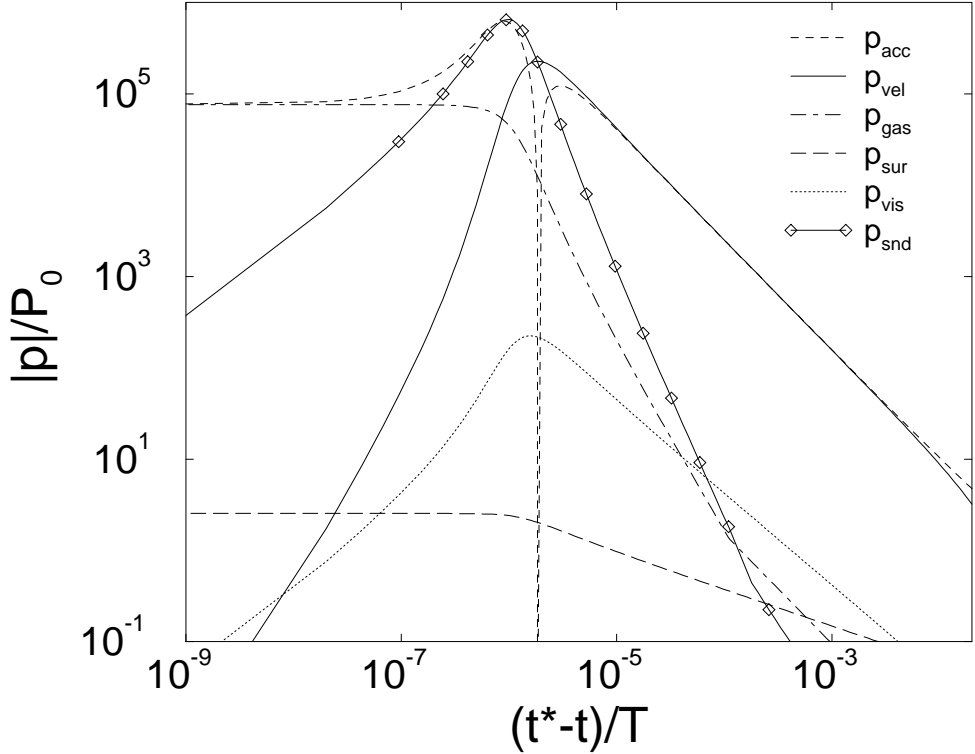,width=13cm,angle=-90.}}
\end{picture}
\caption[]{Relevant pressure
contributions according to the complete RP equation on a
log-log scale before the instant of collapse $t^*$.
The bubble is driven at $P_a=1.4\,atm$ and its ambient radius
is $R_0=4\,\mu m$.
}
\label{rpcolbefore}
\end{figure}

We first take a closer look at the main collapse (after $R_{max}$ has
been reached, interval C in figure~\ref{roft}$b$).
Figure \ref{rpcolbefore} shows the
behaviour of the most important terms in the RP equation
(defined in table \ref{table1}) just prior to the main collapse.
The abscissa displays
the logarithm of the time interval before the collapse time
$t^*$ which is identified by the condition $\dot{R}(t^*)=0$, i.e., the
bubble reaches its minimum radius at $t^*$. The ordinate gives the
logarithms of the absolute values of the various pressure contributions.
As the whole time interval treated in this subsection only comprises
$\approx 0.1\,\mu s$,
and we want to discuss processes as fast as $1\,ns$, we choose the
polytropic exponent in (\ref{vdw}) to be $\kappa=5/3$, the adiabatic value
for argon.
Note that the portions of the graphs for $|t^*-t|\aleq 10^{-7}T$
in figure~\ref{rpcolbefore}, as well as in figures~\ref{roftcol} --
\ref{expsnd} below, represent
time scales on or below the picosecond scale. As hydrodynamics breaks
down here, this part of the computation will only be able to give a
reasonable effective dynamics. We will take care not to draw physical
conclusions from data in this range.
  
%caption8

In a large part of the collapse phase
(figure~\ref{rpcolbefore}) the dynamical terms
$p_{acc}$ and $p_{vel}$ give the dominant contribution; they
compensate each other, so that the dynamics is well described by the
classical Rayleigh collapse
\begineq
R \ddot R + {3\over 2} \dot R^2  = 0.
\label{raycol}
\endeq
This formula complements the quasistatic approximation (\ref{blake}) above.
Equation~\ref{raycol} implies a scaling law for $R(t)$: 
\begineq
R(t)= R_R \left({t^*-t\over T}\right)^{2/5}\, .
\label{scalecol}
\endeq
Here, the oscillation period $T$ is used for non-dimensionalization of the
time coordinate.
The characteristic radius $R_R$ can be estimated from an
energy argument: at $R=R_{max}$, the potential energy of the bubble
is approximately $E_{pot}\sim 4\pi P_0 R_{max}^3/3$, see e.g.\ \cite{sme87}.
Converting this
into kinetic energy of the fluid at $R=R_0$, we get as an estimate for the
bubble wall speed at $R=R_0$
\begineq
\dot{R}\left|_{R=R_0}\right.= -\left({2P_0\over 3\rho_l}\right)^{1/2}
\left({R_{max}\over R_0}\right)^{3/2}\, .
\label{r0dot}
\endeq
Using the time
derivative of (\ref{scalecol}), we find
\begineq
R_R=R_0 \left({5T\left|\dot{R}\left|_{R=R_0}\right.\right|\over 2R_0}
\right)^{2/5}
=\left({25P_0T^2\over 6\rho_l}\right)^{1/5} R_{max}^{3/5}
\approx 14.3\,\mu m\cdot\left(R_{max}\over \mu m\right)^{3/5}\, .
\label{rr}
\endeq
With this $R_R$, (\ref{scalecol}) is compared to the numerical result of
the RP ODE in figure~\ref{roftcol}. Both slope and prefactor are reproduced
excellently, despite the rather crude approximations leading to (\ref{rr}).
The only characteristic value for the Rayleigh collapse is $R_{max}$, which
depends on $P_a$ and (although weakly) on $R_0$. Analytical
expressions for these dependences will be given in Section~\ref{secexp}.

%caption9
\begin{figure}[htb]
\setlength{\unitlength}{1.0cm}
\begin{picture}(12,11)
\put(0.,0.2){\psfig{figure=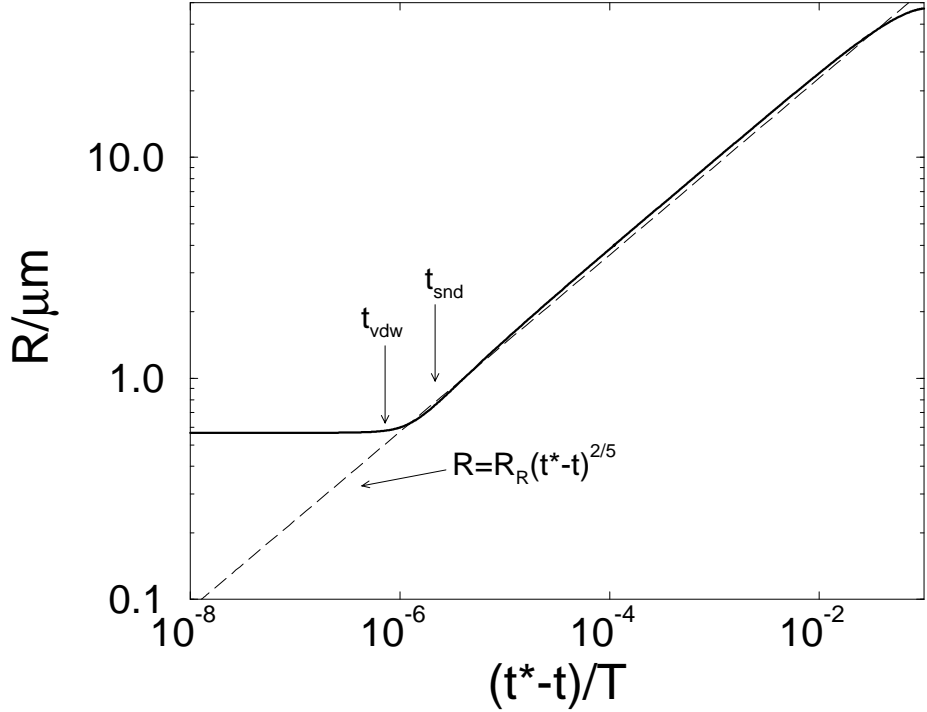,width=13cm,angle=-90.}}
\end{picture}
\caption[]{Collapse dynamics of $R(t)$ for the same parameters as in
figure~\ref{rpcolbefore}. This bubble reaches a maximum radius
$R_{max}\approx 47\,\mu m$ before collapsing.
The theoretically expected Rayleigh scaling law
$R(t)=R_R(t^*-t)^{2/5}$ (dashed) is followed accurately. Also indicated
are the limiting times $t_{vdw}, t_{snd}$ for the validity of the scaling
law.
}
\label{roftcol}
\end{figure}

We now examine the range of validity of (\ref{raycol});
one could worry whether it is justified to neglect $p_{gas}$ and $p_{snd}$
during collapse. For the solution (\ref{scalecol}),
we have $p_{vel}=-p_{acc}\propto (t^*-t)^{-6/5}$, whereas
(as long as $R(t)^3\gg h^3$)
$p_{gas}\propto (t^*-t)^{-2}$ and $p_{snd}\propto(t^*-t)^{-13/5}$ for
$\kappa=5/3$, i.e., the latter two pressure contributions grow stronger
than the dynamical terms as $t\to t^*$. This can also be observed in
figure~\ref{rpcolbefore}, but the absolute value of $p_{gas}$ and $p_{snd}$ is
negligible compared to $p_{vel}, p_{acc}$ except for times very close
to $t^*$. We can compute the range of validity of (\ref{raycol}) by
equating $p_{gas}=p_{vel}$ and $p_{snd}=p_{vel}$, respectively, using
(\ref{scalecol}), (\ref{rr}). 
It turns out that the sound pressure contribution is the first to violate
(\ref{raycol}).
This happens at $t_{snd}$ with
\begin{eqnarray}
(t^*-t_{snd})/T&=&\left({192\rho_l c_l^2\over 25P_0}\right)^{1/7}
{R_0\over c_lT}
\left(1+\alpha_s\right)^{5/7}
\left({R_0\over R_{max}}\right)^{18/7} \nonumber \\
&\approx& 1.0\times 10^{-4} \left(1+\alpha_s\right)^{5/7}
\left({R_0\over R_{max}}\right)^{18/7}
\left({R_0\over \mu m}\right)\, ,
\label{tsnd}
\end{eqnarray}
which agrees with
the numerical result e.g.\ in figure~\ref{rpcolbefore}
(where $R_0=4\,\mu m$ and
$R_{max}\approx 47\,\mu m$). For the approximation (\ref{tsnd}),
$R^3(t_{snd})\gg h^3$ was assumed; $\alpha_s$ is the surface tension
parameter introduced in (\ref{omes}). 
The collapse behaviour changes due to $p_{snd}$ shortly before another
assumption for (\ref{raycol}) breaks down: obviously, $R(t)$ cannot be
smaller than 
the van der Waals hard core $h$. Equating $R(t)=h$ using (\ref{scalecol})
with $h=R_0/8.86$,
we obtain the ``hard core time'' 
\begineq
(t^*-t_{vdw})/T=\left({6\rho_l\over 25P_0}\right)^{1/2}
{1\over T}{h^{5/2}\over R_{max}^{3/2}}
\approx 5.5\times 10^{-6} \left({R_0\over R_{max}}\right)^{3/2}
\left({R_0\over \mu m}\right)\, .
\label{tvdw}
\endeq
At $t\approx t_{vdw}$, the van der Waals hard core
cuts off the scaling behaviour abruptly. However, for typical values of
$R_0\approx 4\,\mu m$, the bubble collapses like an empty cavity
for a time interval from $(t^*-t)\sim 1\,\mu s$ down to
$(t^*-t)\sim 100\,ps$ ($t^*-t\sim 0.03T\dots 3\times 10^{-6}T$).

\subsection{Turnaround and delayed reexpansion (region R)}\label{turn}

\begin{figure}[htb]
\setlength{\unitlength}{1.0cm}
\begin{picture}(12,11)
\put(0.,0.5){\psfig{figure=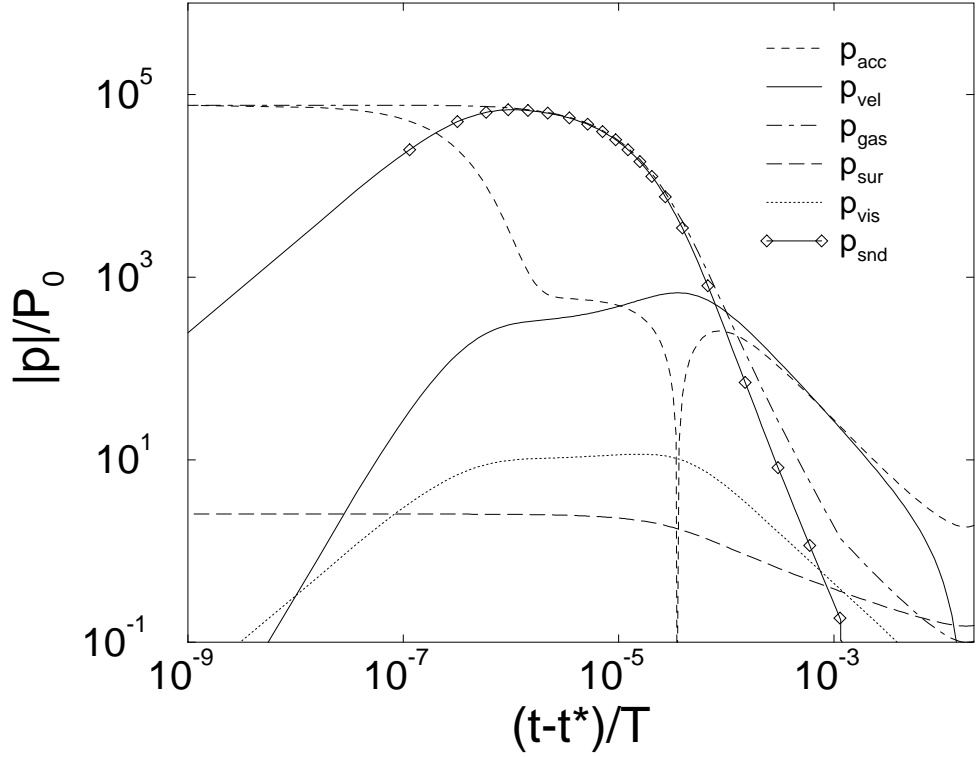,width=13cm,angle=-90.}}
\end{picture}
\caption[]{Important pressure
contributions according to the RP equation on a
log-log scale after the instant of collapse. The same parameters as in
figure~\ref{rpcolbefore} apply.
}
\label{rpcolafter}
\end{figure}

As the gas is compressed to the hard core radius, the collapse is
halted abruptly. L\"ofstedt \etal (1993) have shown that --
in the Hamiltonian limit neglecting $p_{sur}, p_{vis}, p_{snd}$ and the
temporal variation of $p_{ext}$ --
the turnaround time interval of the bubble is approximately
\begin{equation}
\tau_{turn}\equiv\left({R(t^*)\over\ddot{R}(t^*)}\right)^{1/2}
\approx\left[3^\kappa{\rho_l h^2\over P_0(1+\alpha_s)}
\left({h\over R_0}\right)^{3\kappa}
\left({R_{min}-h\over h}\right)^\kappa\right]^{1/2}\, .
\label{loeturn}
\end{equation}
This equation also follows from approximating the RP equation (\ref{rp})
by keeping only the dominant terms in the immediate
vicinity of the collapse, i.e., $p_{acc}$ and $p_{gas}$ (cf.\
figures~\ref{rpcolbefore} and \ref{rpcolafter}):
\begineq
\rho_l R \ddot R = P_0\left(1 + \alpha_s\right)
\left( {R_0^3 - h^3\over R^3(t) - h^3 }\right)^\kappa\, .
\label{rpgasacc}
\endeq
(\ref{rpgasacc}) is a good description of bubble dynamics for a time
interval around the collapse of length $\sim\tau_{turn}$.
 
%caption10

\begin{figure}[htb]
\setlength{\unitlength}{1.0cm}
\begin{picture}(12,10.7)
\put(0.,0.5){\psfig{figure=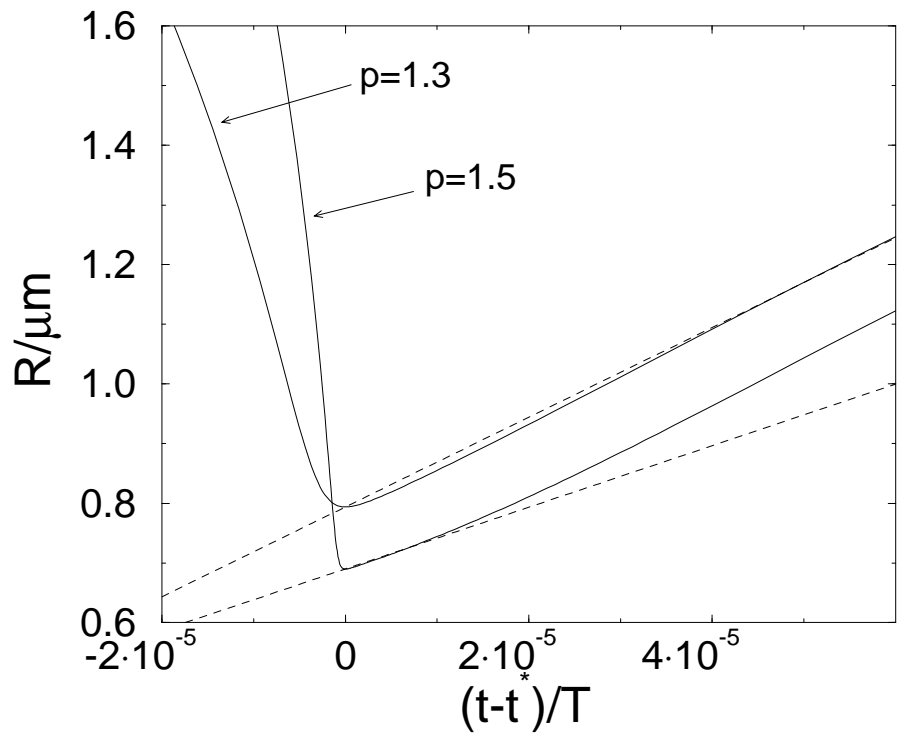,width=13cm,angle=-90.}}
\end{picture}
\caption[]{Bubble dynamics 
for $R_0=5\,\mu m$ and two different driving pressure amplitudes
$p=1.3$ and $p=1.5$ in the vicinity of collapse. The dashed lines give
the linear reexpansion approximation from equation (\ref{soundapp}).
}
\label{expsnd}
\end{figure}

Figure~\ref{rpcolafter} shows the pressure contributions {\em after} $t^*$.
From $(t-t^*)/T\sim 10^{-6}$ to
$(t-t^*)/T\sim 10^{-4}$ (i.e., from $(t-t^*)\sim 30\,ps$ to $\sim 3\,ns$)
the dominant terms in (\ref{rp}) are $p_{gas}$ and
$p_{snd}$, which compensate each other. This means that the energy stored in
the compressed gas is released almost exclusively through emission of
sound waves (cf.~Church 1989) -- it is not converted back to kinetic energy
of the liquid surrounding the bubble.
The corresponding
dynamics shows a relatively low expansion velocity and small acceleration,
keeping a very small bubble radius for a few $ns$ (figure~\ref{expsnd}). 
This time interval of {\em delayed reexpansion} (denoted by R in
figure~\ref{roft}$b$)
is described by
\begineq
0 = p_{gas}(R(t)) + {R(t)\over c_l} {{\mbox d}\over {\mbox d}t}
p_{gas}(R(t))
\label{gassound}
\endeq
with $p_{gas}$ given by (\ref{vdw}).
This ODE has an analytical solution :
\begin{equation}
{c_l\over 3\kappa} (t-t^*) = 
\left[R+{h\over 6} \ln{(R-h)^2\over R^2+h^2+Rh} - {h\over\sqrt{3}}
\arctan {2R+h\over\sqrt{3}h}\right]_{R_{min}}^{R(t)} \, .
\label{soundsol}
\end{equation}
For $(R(t)-R_{min})\ll R_{min}$, i.e., just after the collapse,
this implicit equation can be simplified to yield
\begin{equation}
R(t)\approx R_{min} + {c_l\over 3\kappa} {R_{min}^3-h^3\over R_{min}^3}
(t-t^*) \, .
\label{soundapp}
\end{equation}
This linear expansion law holds for a longer time interval if $R_{min}$
is larger, i.e., for smaller $P_a$. Its validity is demonstrated in
figure~\ref{expsnd}. Note that although the turnaround time $\tau_{turn}$
becomes smaller for decreasing $R_{min}-h$, the velocity of the bubble
immediately after collapse is actually
{\em smaller} because of the larger energy losses through acoustic radiation.
The strongly asymmetric shape of $R(t)$ around $t^*$ has also been
observed in experimental measurements of bubble dynamics, e.g.\ by
\cite{bar92}, Tian \etal (1996), \cite{wen97}, and \cite{mat97}.

%caption11

After the delayed reexpansion phase, the bubble wall gains speed and enters
another short time interval around $(t-t^*)\approx 10^{-3}T$ well
described by Rayleigh's equation (\ref{raycol}) with
$R(t)\propto (t-t^*)^{2/5}$ as the bubble expands.
At $(t-t^*)\approx 10^{-2}T$ it
enters the phase of subsequent afterbounces.

\subsection{Afterbounces: a parametric resonance (region AB)}\label{secafter}

\begin{figure}[htb]
\setlength{\unitlength}{1.0cm}
\begin{picture}(12,11.3)
\put(0.,0.2){\psfig{figure=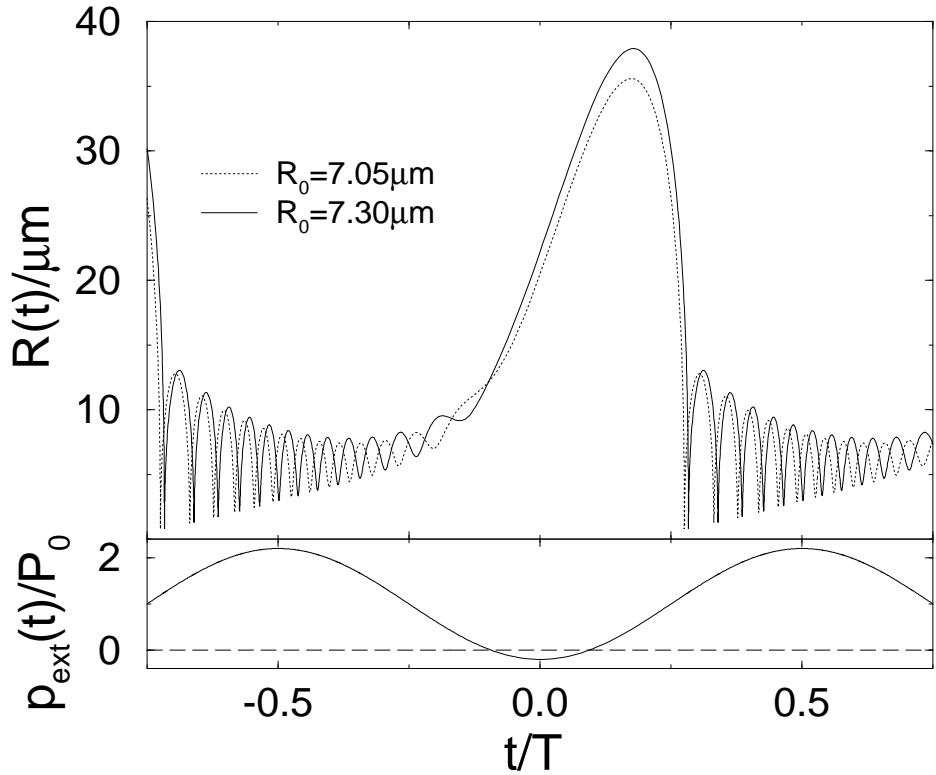,width=13cm,angle=-90.}}
\end{picture}
\caption[]{Two time series $R(t)$ for $P_a=1.2\,atm$ and $R_0=7.05\,\mu m$
(dashed), $R_0=7.30\,\mu m$ (solid).
The first ambient radius corresponds to a resonance minimum in
$R_{max}/R_0$,
the second to a maximum. Note that the afterbounces at the beginning of the
main expansion are just one half-cycle out of phase.
}
\label{minmax}
\end{figure}

The discussion of the afterbounce interval (AB in figure~\ref{roft}$b$)
is intimately connected to
the explanation of the wiggly structure of various dynamically computed
terms, like the expansion ratio (figure~\ref{p4}$b$) or the diffusive
equilibrium lines in Fig \ref{total}.
Obviously, as the RP equation (\ref{rp}) describes a driven oscillator,
the maxima in the expansion ratio represent parameter values of resonant
driving. Figure~\ref{minmax} clarifies the character of these resonances. It
shows two time series of the bubble radius $R(t)$ at values of $R_0$
corresponding to a relative maximum and a relative minimum of $R_{max}/R_0$,
respectively.
A large or small expansion ratio results from the phase of the afterbounces
at the time when $p_{ext}$ becomes negative, i.e., when the external forces
start the rapid expansion: for the bubble with the large expansion
ratio, the last afterbounce ``fits'' into the expansion, which is enhanced.
For the other bubble, growth is inhibited as the last afterbounce collapse
interferes with the expanding external force.

%caption12

The afterbounce oscillations show relatively small amplitude, and it is
therefore possible to linearize the RP equation in this region of the
driving cycle. Moreover, sound radiation and viscosity contributions
are negligible. For simplicity, we also neglect $p_{sur}$ for the moment.
In order to separate the time scale $1/\om$ of the driving
from the much shorter time scale of the afterbounces, which is
$\sim 1/\om_e$, we
use the ansatz (cf.\ e.g.\ Hinch~1991)
\begineq
R(t)=\widetilde{R}_0(\tau)(1+y(t))
\label{separation}
\endeq
with small $y(t)$ and a slowly varying function $\widetilde{R}_0(\tau)$
which is
to be determined;
$\tau=\epsilon t, \epsilon=\om/\om_{e}\ll 1$. 
To leading orders in $y$ and $\epsilon$, equation~(\ref{rp}) is transformed
into
\begineq
\widetilde{R}_0^2\ddot{y} = -\om_e^2{R_0^5\over\widetilde{R}_0^3}y +
{P_0\over\rho_l}\left({R_0^3\over\widetilde{R}_0^3}-(1-p\cos\om_{e} \tau)
\right)
\label{linrp}
\endeq
Requiring the slowly varying (secular) term on the right-hand side 
to vanish, we have to choose
\begineq
\widetilde{R}_0(\tau)=R_0/ \left(1-p \cos(\om_{e} \tau)\right)^{1/3}
=R_0/ \left(1-p \cos(\om t)\right)^{1/3}.
\label{r0tilde}
\endeq
With this definition, (\ref{linrp}) results in a Hill equation:
\begineq
\ddot{y}+\om_e^2(1-p\cos\om t)^{5/3} y = 0.
\label{hill}
\endeq
Because of the separation of time scales $\om_e\gg\om$ this equation
represents
a harmonic oscillator with slowly varying eigenfrequency, i.e., the afterbounce
frequency $\om_{ab}=\om_e (1-p\cos \om t)^{5/6}$. For this system,
$E(\om_{ab})/\om_{ab}$ (with $E=\la y^2\ra \om_{ab}^2/2$
being the oscillator energy)
is an adiabatic invariant (see Hinch 1991), i.e., 
\begineq
\la y^2\ra(1-p\cos \om t)^{5/6}=const. \, ,
\label{adinv}
\endeq
where the mean $\la\cdot\ra$ is an average over the fast time scale.
Note that in the time interval $\pi/2\aleq\om t\aleq 3\pi/2$ of afterbounces
$(1-p\cos\om t)>0$. 
It follows that the amplitude of the afterbounces changes as
$\tilde{R}_0y \propto (1-p\cos \om t)^{-3/4}$.

The resonance structure of (\ref{hill}) still cannot be evaluated
analytically. Yet
the parametric driving of (\ref{hill}) has a very similar shape to the
cosine driving of a Mathieu equation. We can therefore further approximate
(\ref{hill}) by choosing suitable constants $Q_1,Q_2$, where we require
\begineq
Q_1-Q_2 \cos(\om t) = \left(1-p \cos(\om t)\right)^{5/3}
\quad {\mbox{for}}\quad \om t={\pi\over 2},\pi \, ,
\label{equal}
\endeq
i.e., $Q_1=1$, $Q_2=(1+p)^{5/3}-1$.
The errors in this approximation are only a few percent
in the time interval $\sim[\pi/2,3\pi/2]$ of afterbounces
we focus on. As an analytically accessible approximation to
the afterbounce dynamics of (\ref{rp}) we have thus the Mathieu equation 
\begineq
y''+4{\om_e^2\over\om^2}\left(1-\left[(1+p)^{5/3}-1\right]
\cos 2\hat{x}\right) y = 0
\label{mathieu}
\endeq
with dimensionless time $\hat{x}=\om t/2$; the primes denote derivatives
with respect to $\hat{x}$.

The contribution of surface tension may be included if $\alpha_s\ll 1$ to
yield a refinement of
(\ref{mathieu}):
\begineq
y''+4{\om_e^2\over\om^2}
\left(\left(1+{2\over 3}\alpha_s\right)
-\left[(1+p)^{5/3}-1+{2\over 3}\alpha_s\left(1+2p-(1+p)^{5/3}\right)
\right]
\cos 2\hat{x}\right) y = 0 \, ,
\label{mathieusurf}
\endeq
with the factor $\alpha_s$ from (\ref{omes}). Note that a simple substitution
$\om_e\to\om_s$ does not cover all first-order effects of $\alpha_s$.

For certain parameter combinations, equation~(\ref{mathieusurf}) shows
parametrically stable or unstable solutions. 
Because $\om_e/\om\gg 1$, the best analytical approximation to these
characteristic values is given by
the asymptotic series (Abramowitz \& Stegun 1972)
\begineq
b = \nu\sqrt{s} - {\nu^2+1\over 8}-{\nu^3+3\nu\over2^6\sqrt{s}}
\mp \dots
\label{asyseries}
\endeq
with $b = 4{\om_e^2\over\om^2}[(1-2\alpha_s/3)(1+p)^{5/3}+
4\alpha_s/3\cdot(1+p)],\;
s=8{\om_e^2\over\om^2}[(1-2\alpha_s/3)(1+p)^{5/3}
+4\alpha_s/3\cdot p-1+2\alpha_s/3]$, and
$\nu=2k_M\pm1$, where the sign distinguishes even from odd Mathieu solutions.
$k_M$ is the order of the Mathieu resonance, corresponding to the number of
afterbounces in the RP equation (see below).
We take here only the leading term on the right-hand side 
of (\ref{asyseries}) and 
treat the case $k_M\gg1$, so that $\nu\approx2k_M$;
moreover, we only keep terms up to first order in $\alpha_s$.
This yields ambient radii  $R_0^{(k_M)}$ for
which the oscillation shows maximum stability against parametric excitation:
\begin{eqnarray}
R_0^{(k_M)}&=&\left({3P_0\over 2\rho_l\om^2}\right)^{1/2}
{q^{5/3}\over\sqrt{q^{5/3}-1}}
{1\over k_M} +{2\sigma\over 3P_0}\left({q^{5/3}-2q+1\over q^{5/3}-1}+
2{2-q^{2/3}\over q^{2/3}}\right) \nonumber \\
&\approx& 74.0\,\mu m \cdot {q^{5/3}\over\sqrt{q^{5/3}-1}}{1\over k_M} +
0.487\,\mu m \cdot \left({q^{5/3}-2q+1\over q^{5/3}-1}+2{2-q^{2/3}\over
q^{2/3}}\right) \, .
\label{resradii}
\end{eqnarray}
Here we have abbreviated $q=(p+1)$. Note that the correction term due
to surface tension does not depend on $k_M$.

%caption13
\begin{figure}[htb]
\setlength{\unitlength}{1.0cm}
\begin{picture}(12,11.3)
\put(0.,0.2){\psfig{figure=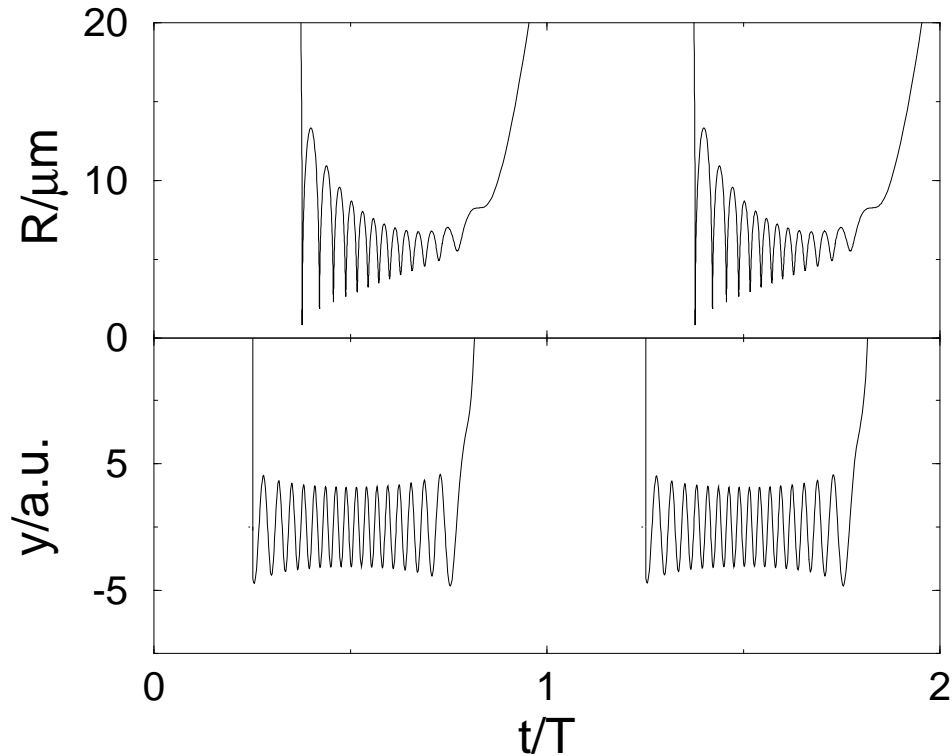,width=13cm,angle=-90.}}
\end{picture}
\caption[]{Comparison of time series from the numerical solution of the
RP equation (\ref{rp}) (upper) and a $T$-periodic solution of
the Mathieu approximation (\ref{mathieusurf})
(lower) for
the same parameters $p=1.5, R_0=6.0\,\mu m$.
The onset of afterbounces is delayed in the full RP dynamics. The correction
factor (\ref{ckexp}) has to be employed. As the Mathieu solution is divergent
(grows exponentially from cycle to cycle), the lower curve was normalized
once per cycle.  
 }
\label{dynmathrp}
\end{figure}

Although the behaviour of the RP oscillator is fairly well described by Mathieu
oscillations in the afterbounce phase (see e.g.\ figure~\ref{dynmathrp}),
it is of course entirely
different during the expansion interval of the cycle. Therefore,
some information about the overall shape of the oscillation must enter
into our analysis. Especially, Mathieu
solutions can be $T$- or $2T$-periodic. RP dynamics in the SL regime, however, 
only allows for $T$-periodic solutions, as the $2T$-periodic Mathieu
solutions would require large negative values for~$y$.
Therefore, every second
resonance of (\ref{mathieusurf}) must be dropped, i.e., the resonance of order
$k_M$ of
(\ref{mathieusurf}) corresponds to resonance number $k=k_M/2$ of (\ref{rp}),
so that the $k^{th}$ resonance radius $R_0^{(k)}$ of (\ref{mathieusurf})
for a dynamics with $k$ afterbounces is obtained
by replacing $k_M$ by $2 k$. 

We must also provide additional information about the length of the afterbounce
interval. Figure~\ref{dynmathrp} shows that this length, which is almost  
independent of $p$ for our Mathieu approximation, is significantly
reduced for increasing $p$ in the case of the RP equation, as the
expansion interval lasts
longer. This is a property of the nonlinear part of the RP cycle (cf. the next
subsection), which cannot
be modelled within the Mathieu approximation.
In the Appendix it is shown that the length of the
afterbounce phase (and therefore the resonance number $k$) has to be
rescaled according to $k\to C(p)\times k$ with $C(p)$ approximately given
by the expansion
$C(p)\approx 0.688-0.548(p-\pi/2)+0.418(p-\pi/2)^2$ (cf.\ (\ref{ckexp})). 

%caption14
\begin{figure}[htb]
\setlength{\unitlength}{1.0cm}
\begin{picture}(12,11.3)
\put(-0.3,0.5){\psfig{figure=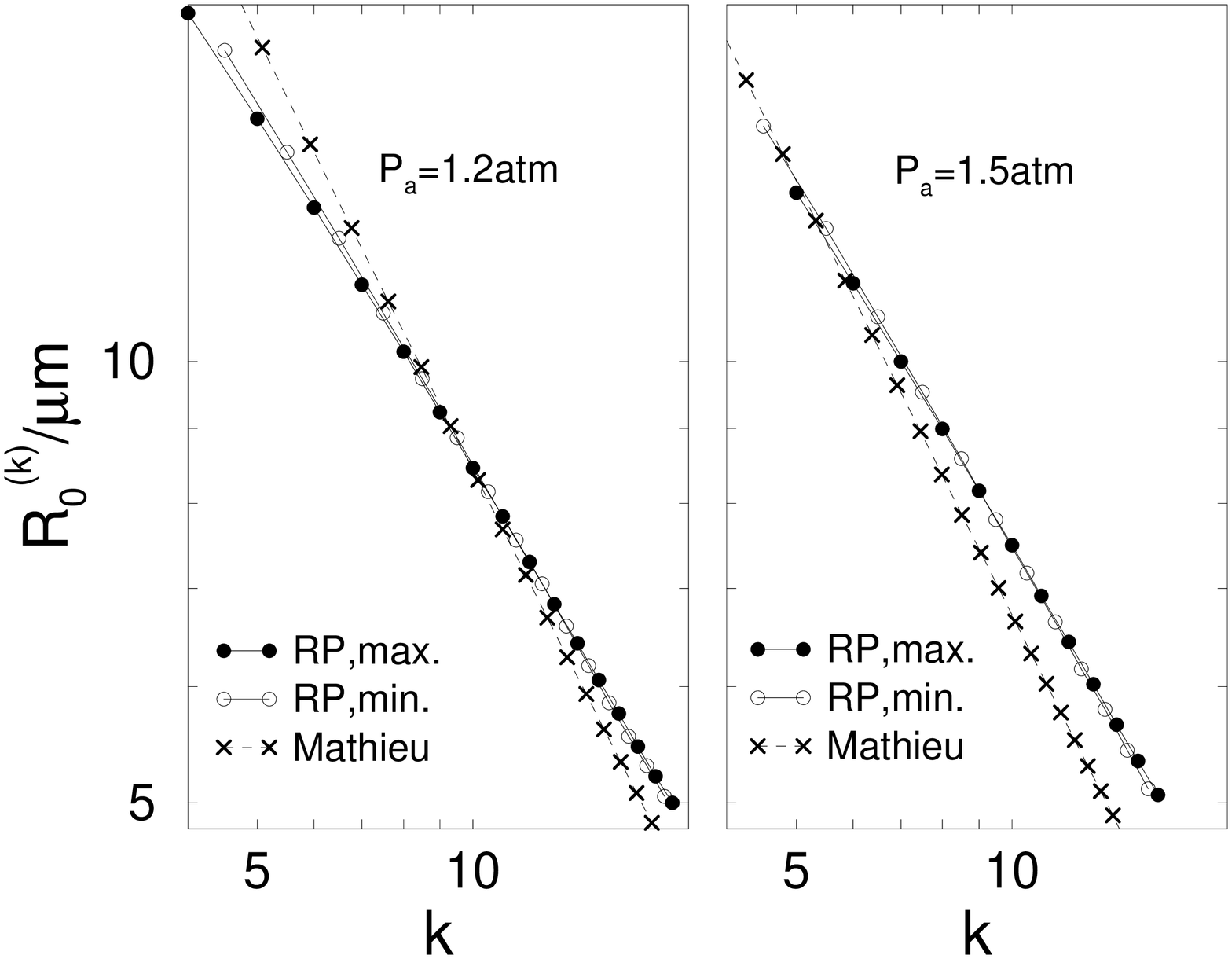,width=13cm,angle=-90.}}
\end{picture}
\caption[]{Ambient radii $R_0^{(k)}$ as a function of resonance order
$k=k_M/2$ from the
full solution of (\ref{rp}) (circles)
and the analytical Mathieu approximation (\ref{resradii}) (crosses);
$P_a=1.2\,atm$ (left) and 1.5atm (right). Note that the
circles represent relative maxima {\em and} minima of $R_{max}$, because
the Mathieu equation admits twice as many resonance values as the RP equation.
The Mathieu solutions were rescaled by the factor $C(p)$ of
equation~(\ref{ckexp}).
}
\label{comprpmath}
\end{figure}

Figure~\ref{comprpmath} presents a comparison of the computed resonance radii
of order (number of afterbounces) $k$ from numerical solutions of (\ref{rp}) --
both for relative maxima and minima of $R_{max} (R_0)$ -- and
from (\ref{mathieusurf}) for driving pressure amplitudes $p=1.2$ and 1.5,
corrected with $C(p)$.
The resonance locations are in good agreement for both pressures,
considering the multitude of approximations they were calculated with.

In the stability maxima marked by $R_0^{(k)}$
the bubble is less excitable by the driving than
bubbles with neighboring $R_0$. Therefore the expansion ratio has a 
local minimum and the average pressure $\la p_{gas}\ra_{4} \propto
1/R_{max}^3$ experiences a local maximum. 
The existence of such wiggles in $R_{max}/R_0$
(and therefore in $\left< p_{gas}\right>_{4}$)
leads to the possibility of multiple equilibria for given experimental
parameters $P_a$ and $p_\infty$: the ambient radius can adjust itself
diffusively to different stable equilibrium values, depending on initial
conditions and/or perturbations. For an analysis of the physical
consequences of multiple
equilibria we refer the reader to the work of \cite{bre96},
Hilgenfeldt \etal (1996) and \cite{cru94b}.

\subsection{Bubble Expansion (regions E$_1$, E$_2$)}\label{secexp}

We wish to be more quantitative about the properties of the expansion phase
now. Despite the small portion of parameter space for SL bubbles, there
are different types of expansion behaviour to be identified depending on
$p$ and $R_0$.
For $p\ageq 1$ and large $R_0\ageq 10\,\mu m$, the gas pressure plays an
important role and balances the dynamical pressure, which is dominated
by $p_{vel}$ for most of the cycle, so that a first approximation to the
dynamics is:
\begineq
\rho_l {3\over 2} \dot R^2 = p_{gas}(R,t)
\label{exp2}
\endeq
With $p_{gas}(R,t)\approx P_0 R_0^3/R^3$ for large $R_0$, this equation yields
a solution for R(t):
\begineq
R(t) = \left[R_-^{5/2}+{5\over 2}\left({2P_0\over 3\rho_l}\right)^{1/2}
R_0^{3/2} (t-t_-)\right]^{2/5} 
\label{sol35}
\endeq
with the starting time of expansion $t_-$ and starting radius $R_-=R(t_-)$.
For radii $R\gg R_-$, (\ref{sol35}) reduces to a
Rayleigh-type expansion law, which gives a scaling relation for the
maximum radius:  
\begineq
R_{max} \propto R_0^{3/5} \,
\label{scale35}
\endeq
if we assume that the length of the expansion interval is independent of $R_0$,
which is a good assumption except for small $R_0$ below the Blake
threshold (\ref{r0trans}). The law (\ref{scale35}) can numerically be confirmed
for large $R_0$, see Hilgenfeldt \etal (1996). Together with (\ref{diffp4rmax})
this yields $\la p_{gas} \ra_{4}\propto R_0^{6/5}$.

\vspace{0.3cm}

\begin{figure}[htb]
\setlength{\unitlength}{1.0cm}
\begin{picture}(12,11)
\put(-0.2,0.){\psfig{figure=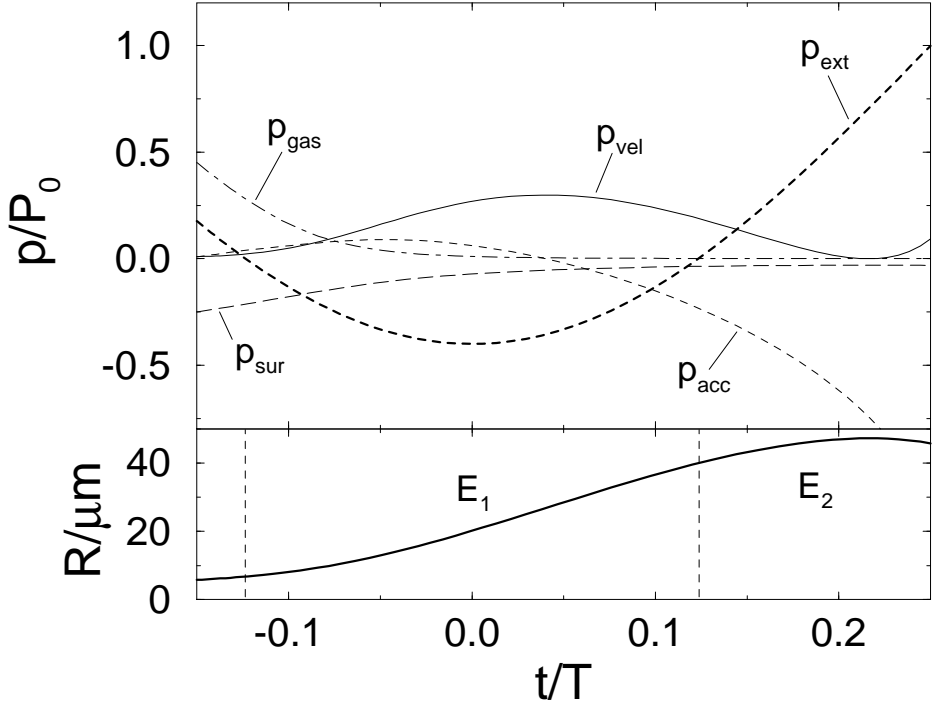,width=13cm,angle=-90.}}
\end{picture}
\caption[]{Upper: important pressure contributions
for $P_a=1.4\,atm$, $R_0=4.0\,\mu m$ during the
expansion phase.
The thick dashed line gives the external pressure
$p_{ext}=P_0-P_a\cos\om t$. In the time interval E$_1$, it is primarily \
balanced by $p_{vel}$ (solid), in E$_2$ by $p_{acc}$ (dashed).
Lower: bubble radius expansion dynamics for the same parameters.
}
\label{exp1440}
\end{figure}

For higher driving pressure amplitudes or smaller $R_0$,
i.e., in most of the SL parameter region, the approximation (\ref{exp2})
is too crude. Instead, one has to take into account the 
balance of the dynamical pressures
$p_{acc}, p_{vel}$ and the external pressure $p_{ext}$
\begineq
\rho_l \left( R \ddot R + {3\over 2} \dot R^2 \right)  =
P_0 (p\cos \om t - 1)\, ,
\label{exp1}
\endeq
as can be seen e.g.\ in
figure~\ref{exp1440}.

%caption15

In the work of L\"ofstedt \etal (1993),
the left-hand side of (\ref{exp1}) has been approximated using a power
series for $R(t)$.
This leads to a
bubble expansion which is linear in time, with a velocity proportional to
$\sqrt{p-1}$. The first nonlinear corrections are of fourth order in~$t$.

%caption16
\begin{figure}[htb]
\setlength{\unitlength}{1.0cm}
\begin{picture}(12,11)
\put(-0.2,0.){\psfig{figure=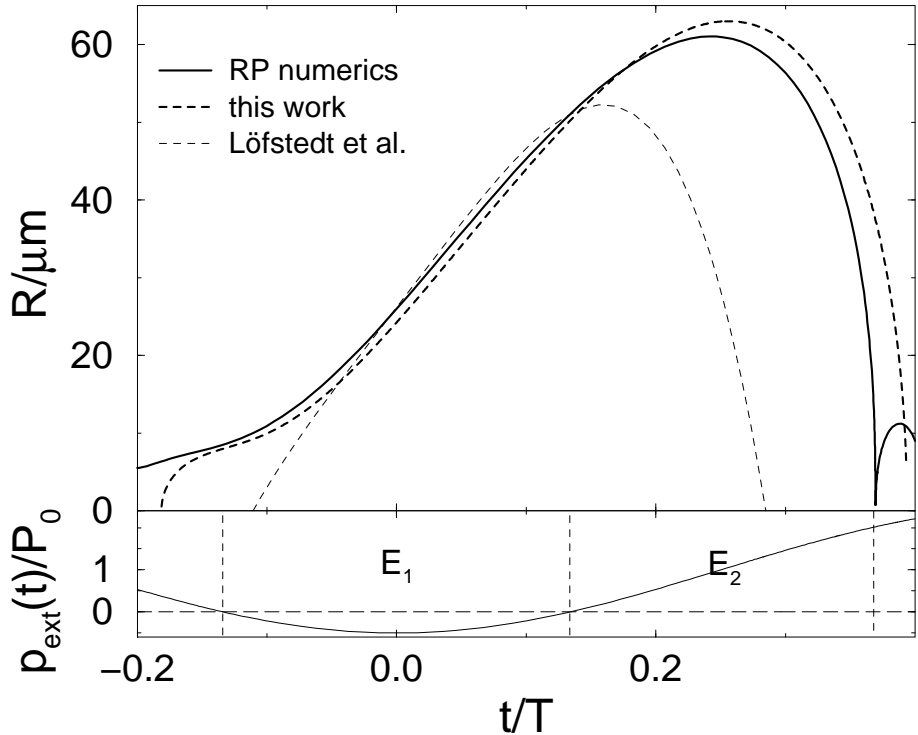,width=13cm,angle=-90.}}
\end{picture}
\caption[]{Comparison of RP dynamics (solid) and theoretical approximation
(dashed) of bubble expansion dynamics for $P_a=1.5\,atm$, $R_0=5.0\,\mu m$.
The theoretical solution becomes complex left and right of its zeros.
Also shown is the suggested theoretical solution (thin dashed line) from
equation~(29) of
L\"ofstedt \etal (1993), where $R(t=0)$ was matched to the value
of the RP dynamics.
}
\label{expana}
\end{figure}

However, an expansion like this turns out to be a series
which is not well-controlled. Especially, no reliable values for the time
$t_{max}$
and value $R_{max}$ of the maximal radius can be derived (cf.\
figure~\ref{expana}). We therefore follow
a different ansatz: consider the dynamical terms $p_{acc}$ and $p_{vel}$ for
a typical bubble expansion (figure~\ref{exp1440}). During the first,
almost linear part $|p_{vel}|\gg |p_{acc}|$, whereas when the maximum is
approached and the
bubble decelerates, $|p_{acc}|\gg |p_{vel}|$. This suggests a division of
the expansion interval into two parts (denoted by E$_1$ and E$_2$ in
figures~\ref{roft}($b$),\,\ref{exp1440}, and \ref{expana}).
Observing that the combination
$R\ddot{R}+\dot{R}^2$ is just the second derivative of $R^2/2$, (\ref{exp1})
can be approximated with good accuracy by
\begin{subeqnarray}
{d^2\over dx^2}R^2  &=& {4\over 9}R_{res}^2(p\cos x -1)
+ {\cal O}\left(R {d^2\over dx^2} R\right)
\quad {\mbox{for}}
\quad x_-\leq x\leq x_+ \, , \label{expleft}
\\
{d^2\over dx^2}R^2  &=& {2\over 3}R_{res}^2(p\cos x -1)
+ {\cal O}\left(\Big({d\over dx} R\Big)^2\right)
\quad {\mbox{for}} \quad
x_+ \leq x\leq x_m \, .
\label{expright}
\end{subeqnarray}
Here we have introduced the dimensionless time $x\equiv \om t$ and
the linear bubble resonance radius $R_{res}$ from (\ref{rres});
$x_m$ is given by $R(x_{m})=R_{max}$.
The rational prefactors on the right-hand side make sure that the dominant
terms $p_{vel}$ in (\ref{expleft}$a$) and $p_{acc}$ in (\ref{expright}$b$) are
correctly represented, while the other terms gives contributions of the
indicated order in each case. 
The starting time $x_-=x_-(p)$ and the transition time
$x_+=x_+(p)$ between both solutions are given by the zeros of $p_{ext}$, i.e.,
$x_+(p)=-x_-(p)=\arccos{1/p}$. Equations (\ref{expleft}$a,b$) can
be integrated analytically requiring continuity and differentiability at $x_+$
for the overall solution. To complete the problem, initial conditions at
$x_-$ have to be imposed: we set $R_-=R(x_-)=\zeta R_0$ with a parameter
$\zeta\sim 1$ whose value is not crucial for the shape of the solution.
An estimate of $\zeta$ can be computed from algebraic equations, but not
in an explicit form. $\zeta$ lies between $1.2$ and $2.0$ for typical $R_0$ of
SL bubbles; for simplicity, we choose $\zeta=1.6$ in all calculations.
For the initial velocity, we
observe that $x_-$ marks the transition from the afterbounce regime, where
the bubble essentially oscillates with its eigenfrequency, to the expansion
regime, where the governing time scale is the driving period $T$. Therefore,
we set $R'_-=(dR/dx)(x_-)=R_-$, corresponding to
$\dot{R}_-=\om R_-$ in dimensional terms.

\begin{figure}[htb]
\setlength{\unitlength}{1.0cm}
\begin{picture}(12,11)
\put(-0.3,0.2){\psfig{figure=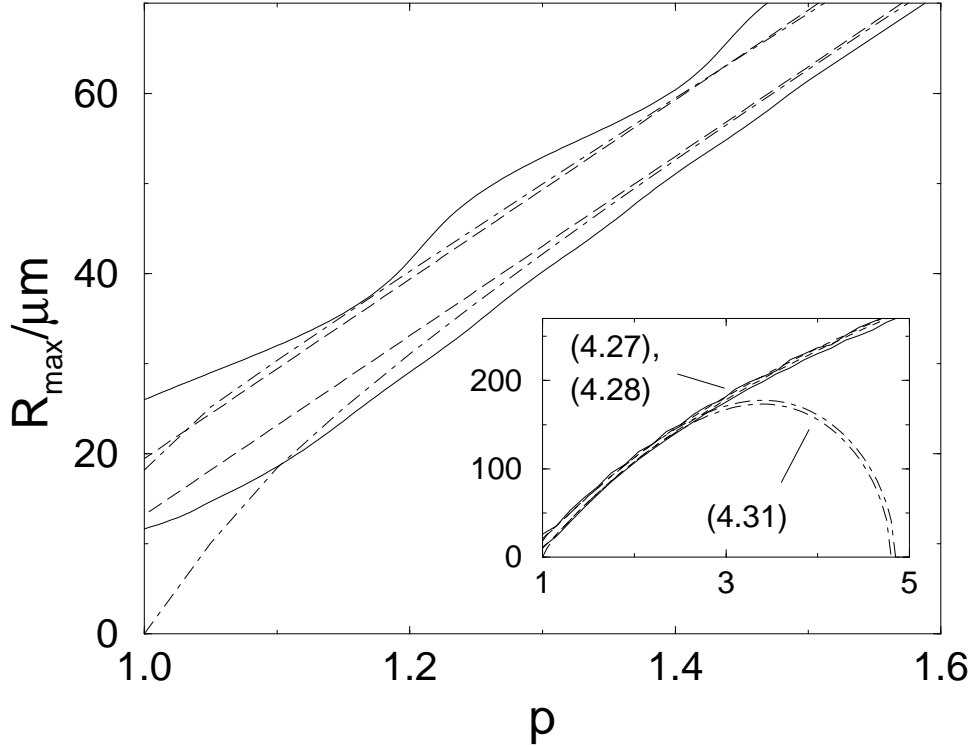,width=13cm,angle=-90.}}
\end{picture}
\caption[]{Comparison of direct numerical results of $R_{max}$ (solid)
for $R_0=5\,\mu m$ (lower) and $R_0=9\,\mu m$ (upper)
in the $p$ range relevant for SL with
the approximate law (\ref{rmaxapprox}) (dot-dashed) and the simpler
formula (\ref{practicalrmax}) (long dashed).
The inset shows that the full theory 
(dashed) according
to (\ref{rmaxana}),\,(\ref{xmaxana}) for $R_{max}(p)$ for the same two $R_0$
gives excellent agreement with the direct RP computation 
(solid)
for both the SL range and the cavitation field
regime $p\sim2-5$. Even the approximation (\ref{rmaxapprox}) yields good
results up to $p\sim 3$ (dot-dashed).
}
\label{rmaxtheo}
\end{figure}

Figure \ref{expana} shows that the shape of the
expansion as well as time and value of the maximum are reproduced
satisfactorily. From the solutions of (\ref{expleft}$a$) in E$_1$ and
(\ref{expright}$b$) in $E_2$ one obtains
a system of equations for $R_{max}$ and $x_{m}$:
\begin{eqnarray}
R_{max}^2 &=& R_-^2 \left(1+2(x_m+x_+)\right) \nonumber \\
&+& {2\over 3}R_{res}^2\left[1-p\cos x_m -{1\over 2}(x_m^2-x_+^2)+
{1\over 3}(p\sin x_+-x_+)(x_m+3x_+)\right],
\label{rmaxana}
\end{eqnarray}
\begin{equation}
p\sin x_m-x_m +{1\over 3}(p\sin x_+-x_+)+{3R_-^2\over R_{res}^2} = 0 \, .
\label{xmaxana}
\end{equation}
Note that (\ref{rmaxana}), (\ref{xmaxana}) give the position and height of
the radius maximum {\em without} any freely adjustable parameters. 
The inset of figure~\ref{rmaxtheo} shows the maximum radii obtained with these
formulas for
$R_0=5\,\mu m $ and $9\,\mu m$ and $p=1-5$ together with results from a
complete RP computation. Apart from the resonance wiggles
(cf.\ Section \ref{secafter}), the curves are very well reproduced
both within the $p$ regime for SL bubbles and for the much higher
pressure amplitudes of cavitation field experiments.

%caption17

\begin{figure}[htb]
\setlength{\unitlength}{1.0cm}
\begin{picture}(12,10.5)
\put(0.,0.5){\psfig{figure=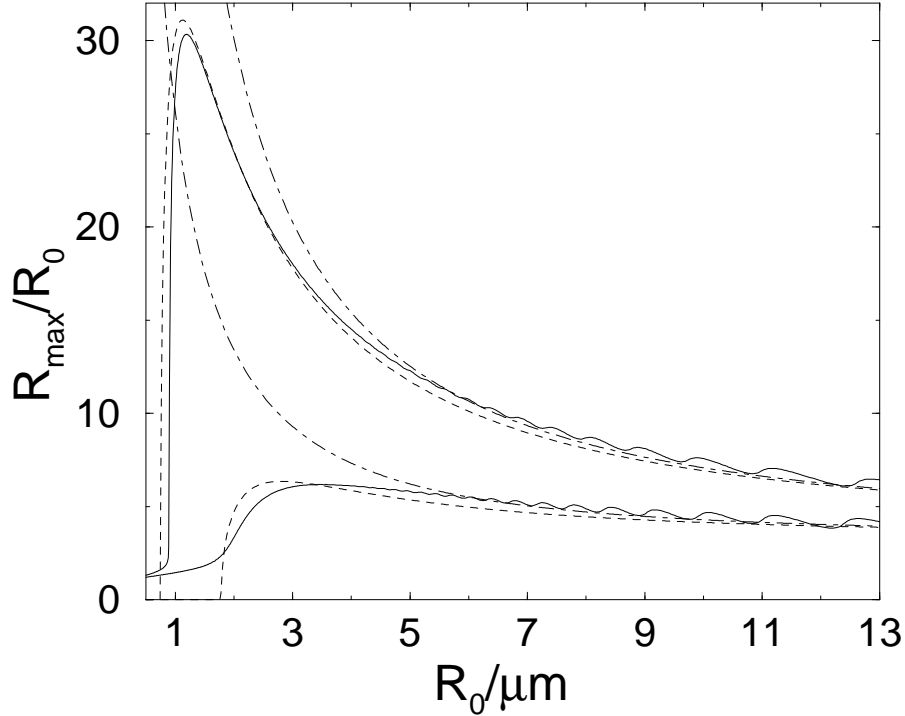,width=13cm,angle=-90.}}
\end{picture}
\caption[]{Direct numerical computation (solid) and
theory (dot-dashed) according to
(\ref{rmaxapprox}) for the expansion ratio $(R_{max}/R_0)(R_0)$ at
$p=1.2$ (lower) and 1.5 (upper). The dashed curve from (\ref{rmaxapproxsurf})
takes surface tension into account and is able to reproduce the numerical
graphs for almost all $R_0$.
}
\label{rmaxr0theo}
\end{figure}

Equation (\ref{xmaxana}) is transcendental, and $R_{max}$
and $x_m$ do not have simple analytical representations. One can, however,
derive simple expressions in several limiting cases:
if $p\gg 1$, we obtain after a lengthy calculation
\begin{equation}
R_{max}\approx\sqrt{2/3}R_{res}(Fp-G)^{1/2}
\label{rmaxlargep}
\end{equation}
with constants $F=1+5\pi/6=3.618\dots,\, G=19\pi^2/24-1=6.813\dots$ .
This formula is a good
approximation only if $p\ageq 5$.
For $p\ageq 1$, i.e., the case of interest for
sonoluminescent bubbles, we can expand $x_m$ around $x=\pi/2$. Moreover,
as $R_{res}\gg R_0$ for SL driving frequencies, we can also neglect the last
term on the right-hand side of (\ref{xmaxana}). To leading order
in ($x_m-\pi/2$), (\ref{xmaxana}) then becomes
\begin{equation}
x_m = p + {1\over 3}(p\sin x_+-x_+) \, ;
\label{xmapprox}
\end{equation}
remember that $x_+=x_+(p)=\arccos(1/p)$.
For $p\in[1.0,1.5]$, the second term of the right-hand side of this 
equation is never
greater than 0.185$p$ in absolute magnitude, so that $x_m = p$ is a
good approximation. Inserting into (\ref{rmaxana}) gives
\begin{subeqnarray}
R_{max}^2 &=& f(p) R_0^2 + g(p) R_{res}^2 \quad {\mbox{with}} \\
f(p) &=& \zeta^2 \left(1+2(p+x_+)\right)\, , \label{rmaxapprox}
\\
g(p) &=& {2\over 3}\left[1-p\cos p -{1\over 2}(p^2-x_+^2)+
{1\over 3}(p\sin x_+-x_+)(p+3x_+)\right] \, .
\end{subeqnarray}
The second term in (\ref{rmaxapprox}a) is much greater than the first;
thus, it is {\em not} primarily $R_0$ which determines
$R_{max}$, but the resonance radius $R_{res}$. With $R_{res}\propto1/\om$,
we see that the expansion ratio is
(at constant $p$ and $R_0$) roughly inversely proportional to the driving
frequency, i.e., upscaling of SL collapses can be achieved by lowering $\om$,
which was also seen in experiment by R.~E.~Apfel (private communication, 1996).
In the same way, a higher ambient pressure $P_0$ (while keeping $p=P_a/P_0$
constant) will lead to higher expansion ratios because of the dependence
of $R_{res}$ on $P_0$ (see (\ref{rres})).
A further simplification of
(\ref{rmaxapprox}) can be obtained from a stringent expansion in $(p-\pi/2)$
and $R_0$, which yields to leading order the simple result
\begineq
{R_{max}\over\mu m}=67.2+0.112\left(R_0\over\mu m\right)^2+99.5(p-\pi/2)
+{\cal O}\left((p-\pi/2)^2\right)\, .
\label{practicalrmax}
\endeq
Figure~\ref{rmaxtheo} shows the very good agreement of (\ref{rmaxapprox}) and
(\ref{practicalrmax}) with
full RP dynamics for several $R_0$ over the whole range of pressures in
SL experiments. The approximation breaks down only at $p\sim 3$, where 
$x_m\approx p$ is no longer valid, see inset of figure~\ref{rmaxtheo}.
The expansion ratio is also accurately reproduced
for moderate or large $R_0$ by this formula, as seen in
figure~\ref{rmaxr0theo}. The deviations for small $R_0$ are due to neglecting
$p_{sur}$, which becomes the dominant influence as $R_0$ approaches
$R_0^{tr}$.

%caption18

One would therefore like to include the effects of surface tension into 
(\ref{rmaxapprox}). We make the following ansatz:
instead of (\ref{expleft}$a,b$), we write
\begin{subeqnarray}
{d^2\over dx^2}R^2  &=& {4\over 9}R_{res}^2\left(p\cos x -
\left(1+{\alpha_s\over K(p)}\right)\right) \quad {\mbox{for}}
\quad x_-\leq x\leq x_+ \, , \label{expleftsurf}
\\
{d^2\over dx^2}R^2  &=& {2\over 3}R_{res}^2\left(p\cos x -
\left(1+{\alpha_s\over K(p)}\right)\right)\quad {\mbox{for}} 
\quad x_+ \leq x\leq x_m \, .
\label{exprightsurf}
\end{subeqnarray}
This models the influence of $p_{sur}$ by an average pressure
contribution of $P_0\alpha_s/K(p)$, where $K$ is taken to be
independent of $R_0$.
Expanding $x_m$ again
around $\pi/2$, we get 
\begineq
R_{max}^2 = f(p) R_0^2 + \left[g(p)+{2\over 3}{\alpha_s\over K(p)}\left(
{1\over 2}(p^2+x_+^2)+
{1\over 3}px_+\right)\right] 
R_{res}^2 \, 
\label{rmaxapproxsurf}
\endeq
with $f(p)$, $g(p)$ from (\ref{rmaxapprox}$b,c$).
With this expression, $(R_{max}/R_0)(R_0)$ shows a global maximum at
\begineq
R_0^c(p)={3\sigma\over P_0K(p)}{{1\over 2}(p^2+x_+^2)+
{1\over 3}px_+\over
g(p)}
\label{r0csurf}
\endeq
For large enough $p$, we can equate (\ref{r0csurf}) and (\ref{r0tlargep}),
because $R_0^{tr}$ and $R_0^c$ are very close then. This gives an estimate
for $K(p)$:
\begineq
K(p)={9\over4}\sqrt{3}(p-1){{1\over 2}(p^2+x_+^2)+
{1\over 3}px_+\over
g(p)}
\label{ksurf}
\endeq
In the regime of SL driving pressures ($p=1.2\dots 1.5$) $K(p)$ depends only
weakly on $p$; its value lies between $7.5$ and $9.4$.  
The ansatz (\ref{expleftsurf}$a,b$)
proves very successful for
the description of $(R_{max}/R_0)(R_0)$ over the whole range of relevant
$R_0$, as can be seen from figure~\ref{rmaxr0theo}.

\begin{figure}[htb]
\setlength{\unitlength}{1.0cm}
\begin{picture}(12,11)
\put(0.5,0.2){\psfig{figure=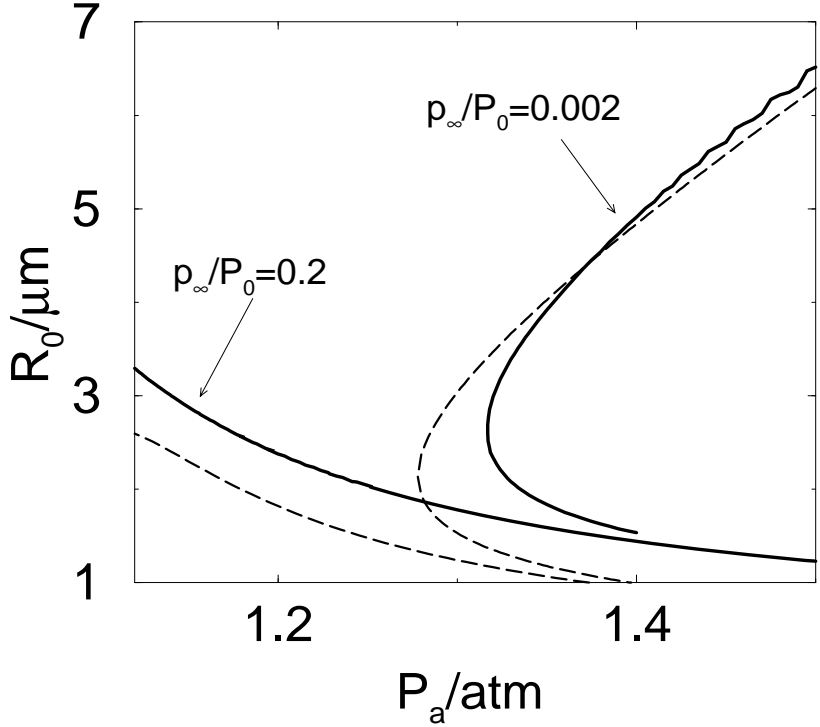,width=13cm,angle=-90.}}
\end{picture}
\caption[]{Diffusive equilibrium lines from figure~\ref{total} (solid) 
and their approximation using (\ref{rmaxapproxsurf}) (dashed) near 
the parameter domain of sonoluminescing bubbles. 
}
\label{diffeqsurf}
\end{figure}

From (\ref{rmaxapproxsurf}), we obtain expansion ratios and, using
(\ref{diffeq}) and (\ref{diffp4rmax})
for given gas concentration $p_\infty/P_0$, an approximation for the
location $R_0(p)$ of diffusive equilibria can be computed.
Figure~\ref{diffeqsurf} shows that both the stable and the unstable branches 
of the equilibrium curves (taken from figure~\ref{total}) are
reproduced satisfactorily for both high and low gas concentrations.

%caption19

When, starting on the stable branch, $p$ is
lowered, $R_{max}/R_0$ becomes smaller and, by (\ref{diffp4rmax}),
$\la p_{gas}\ra_{4}$ becomes larger.
The corresponding equilibrium ambient
radii $R_0$ shrink. Eventually, the minimum of
$\la p_{gas}\ra_{4}/P_0$ becomes larger than $p_\infty/P_0$ (see
figure~\ref{p4}) and no $R_0$ can fulfill the equilibrium condition. This
situation corresponds to the turning point of the diffusive equilibria in
figure~\ref{total} and figure~\ref{diffeqsurf}.

\section{Role of surface tension and liquid viscosity}\label{secsurfvis}

The previous sections have provided a detailed analysis  of
the dynamics of
SL bubbles. How will these results change if we introduce a different fluid
with different surface tension $\sigma$ and/or fluid viscosity
$\eta_l$?

\begin{figure}[htb]
\setlength{\unitlength}{1.0cm}
\begin{picture}(12,11)
\put(-0.2,0.5){\psfig{figure=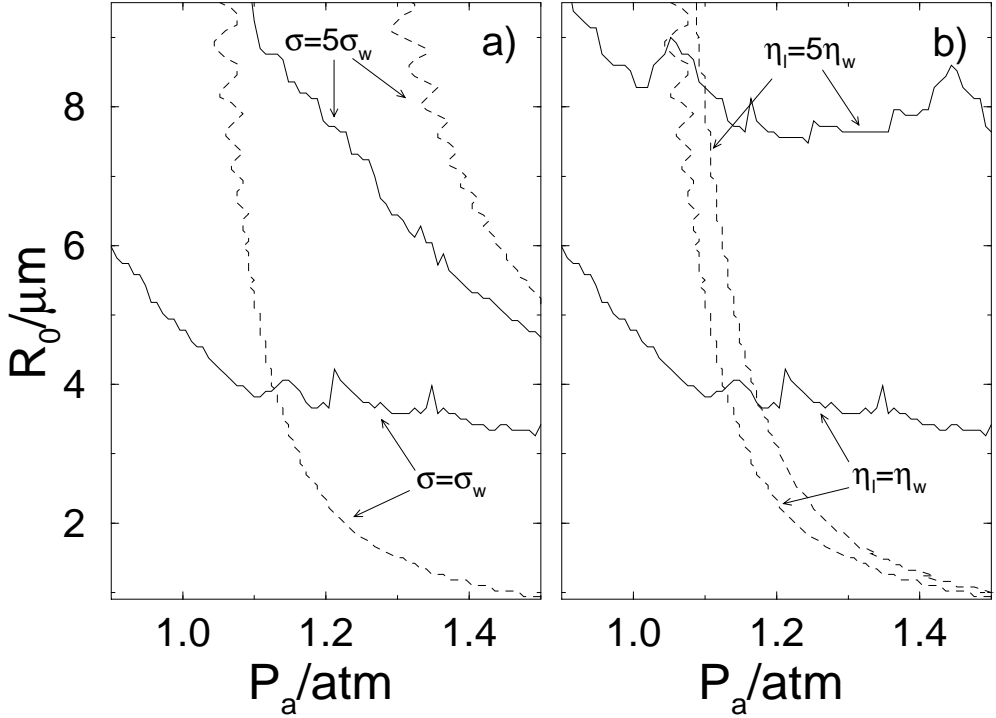,width=13cm,angle=-90.}}
\end{picture}
\caption[]{Parametric stability thresholds (solid), calculated as elaborated
in Hilgenfeldt \etal (1996), and $|M_{g}|=1$ lines (dashed)
for bubbles of different surface tension ($a$) and viscosity ($b$). All other
parameters were fixed at the values for water. Increasing $\sigma$ makes
the region of potential SL bubbles above the $|M_{g}|=1$ and below the
stability threshold vanish, whereas higher $\eta_l$
enlarges this area. 
}
\label{nusig}
\end{figure}

Surface tension is the crucial parameter for the location of the Blake
threshold in parameter space (cf.\ also L\"ofstedt \etal 1995 or
Akhatov \etal 1997).
The transition from weakly oscillating to
strongly collapsing bubbles and therefore the boundary of the SL region
determined by (\ref{mach}) is entirely controlled by $\sigma$.
If we had
$\sigma\to 0$, bubbles with any $R_0$ would be strongly collapsing, i.e.,
liquids with small surface tension should allow for violent collapses
at smaller $P_a$. On the other hand, in liquids with high $\sigma$ larger
$P_a$ and $R_0$ are required for collapses. Although a larger $\sigma$ has a
stabilizing
effect on the bubble surface, (\ref{r0tlargep}) and
(\ref{r0csurf}) show that
the $|M_{g}|=1$ line overtakes the shape instability threshold
in $P_a$--$R_0$ parameter space (cf.\ Hilgenfeldt \etal 1996),
so that no stable SL should
be possible if $\sigma$ is e.g.\ five times higher than in water.
This is easily confirmed by the numerical solution of the RP equation, see
figure~\ref{nusig}($a$). 

%caption20

\begin{figure}[htb]
\setlength{\unitlength}{1.0cm}
\begin{picture}(12,11)
\put(-0.2,0.5){\psfig{figure=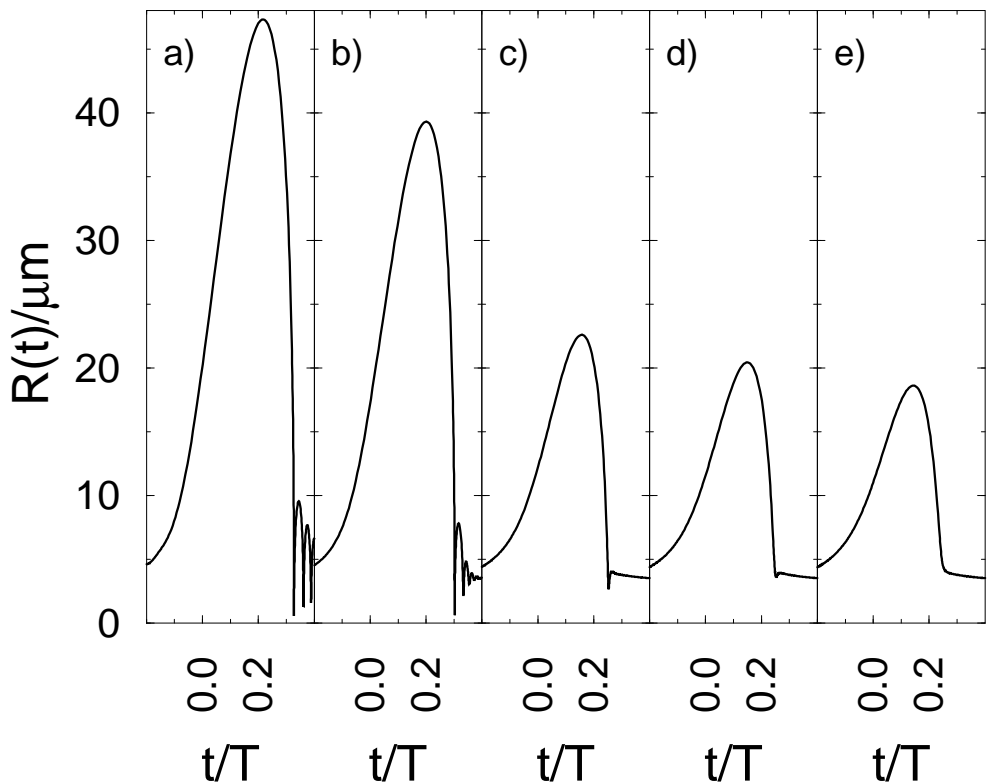,width=13cm,angle=-90.}}
\end{picture}
\caption[]{Bubble dynamics $R(t)$ for different viscosities
$\eta_l/\eta_{\em water}=1$($a$), 10($b$), 35($c$), 40($d$), 45($e$).
$R_0=4\,\mu m$ and $P_a=1.4\,atm$
in all cases. The transition from collapse/afterbounce behaviour to
aperiodic dynamics occurs near $\eta_l/\eta_{\em water}=40$, as predicted from
(\ref{etacrit2}).
}
\label{etac}
\end{figure}

At first sight it seems that fluid viscosity could have been neglected in
(\ref{rp}) right from the start. Apart from a slight
damping effect
during the afterbounce phase, the influence of $\eta_l$ for water
on bubble dynamics is hardly noticeable, 
even a tenfold increase of $\eta_l$ only reduces the maximum radius by
$\approx 10\%$ (figure~\ref{etac}$a$,$b$).
We can, however,
estimate by how much the viscosity would have to be enhanced to have a
significant effect: the damping of a high viscosity liquid should ultimately
prevent the bubble from collapsing violently and therefore it will never
fulfill the energy transfer condition (\ref{mach}). As the collapse is in
fact the first afterbounce minimum, an estimate for this
critical $\eta_l^c$ can be obtained if we demand the afterbounces to be
overdamped. The viscosity $\eta_l$ introduces a
damping term in the linearization of the RP equation (\ref{fulllin}).
It is easy to see that overdamped motion requires
\begineq
{2\eta_l^c\over \rho_lR_0^2}\ageq \om_e\left(1+{\alpha_s\over 3}\right)\, .
\label{etacrit1}
\endeq 
With the definition of $\om_e$, it follows
\begineq
\eta_l^c\ageq \left(1+{\alpha_s\over 3}\right)
\left({3\over 4}\rho_l P_0 R_0^2\right)^{1/2}\, .
\label{etacrit2}
\endeq
With a typical value $R_0=4\mu m$ and keeping $\sigma$, $\rho_l$ at
the values for water,
we obtain that $\eta_l^c\ageq 40\eta_{\em water}$. This is confirmed by
direct computation of (\ref{rp}) using $\eta_l^c$ and strong driving
pressure $P_a=1.4\,atm$, see figure~\ref{etac}($c$)-($e$).
Viscosities in this range can be easily achieved in mixtures of water and
glycerine. For moderate glycerine percentage, the viscosity is not
very different from pure water, but for high glycerine contents it rises
dramatically. The required factor of 40 is (at $10^\circ$C) reached for
$\approx 70\%$ glycerine (weight percentage). Above this percentage, it would
be extremely  difficult to obtain collapses strong enough to ensure energy
transfer and the ignition of SL. Moreover, chemical dissociation reactions
in air cannot take place, which seem to be necessary for SL stability using air
at moderate degassing levels (Lohse \etal 1997).
This may be the reason why \cite{gai90} was not
able to observe stable bubbles above a glycerine percentage of $\approx 60\%$.
The actual threshold for SL should occur at slightly smaller $\eta_l$ than
predicted by (\ref{etacrit2}), because even if the collapse minimum is not
completely damped out, the collapse is already considerably weakened
(figure~\ref{etac}$c$). Also, the threshold should be higher because of the
additional damping effect of thermal dissipation (see Vuong \& Szeri 1996,
Yasui 1995)
which is not included in our approach.

%caption21

Even for smaller $\eta_l\sim\eta_{\em water}$, fluid viscosity is an
important contribution to the damping of bubble surface oscillations,
as was shown in Hilgenfeldt \etal (1996). 
Therefore, a moderate increase in $\eta_l$
does not lead to significant changes in the $R(t)$ dynamics itself
(and therefore in the $|M_{g}|=1$ curve), but
it helps to stabilize bubbles at larger $R_0$. This change affects only
the parametric and afterbounce instabilities (see Hilgenfeldt \etal 1996),
which can
show accumulation effects over several driving periods, but not
the Rayleigh-Taylor instability, which is directly dependent on the
acceleration of the bubble wall and cannot change significantly when the
$R(t)$ dynamics does not. Therefore, the Rayleigh-Taylor process
still cuts off the bubble stability region (cf.\ figure~6  of
Hilgenfeldt \etal 1996)
at $P_a\approx 1.45atm$ almost
independent of viscosity, whereas at smaller $P_a$, much larger bubbles
can be stable if $\eta_l$ is enhanced.
 
In figure~\ref{nusig}($b$), we
show surface stability curves for different fluid viscosities. For high
$\eta_l$, the
region of stably sonoluminescing bubbles between the almost unaffected
$|M_{g}|=1$ curve and the increased stability threshold is considerably
enlarged.
This would probably correspond to a substantial upscaling of SL intensity.
Gaitan's (1990) experimental observation that a moderate
percentage of glycerine helps to establish stable SL bubbles 
supports this conjecture (see also the experimental results of Gaitan \etal
1996 for fluids of different viscosity and surface tension).
Combining our results for $\sigma$ and $\eta_l$, we conclude that
the ideal fluid for violently collapsing, but surface stable
bubbles should have small surface tension and high viscosity.

\section{Conclusions}\label{secconcl}

The analysis of the RP equation presented here
has explained quantitatively quite a lot of features seen in numerical
computations of RP
dynamics. One cycle of oscillation of this
highly nonlinear system can be completely divided into subsections in which its
behaviour can be accurately approximated by analytically integrable equations.
While being in the spirit of L\"ofstedt \etal's (1993) previous analysis,
the present work presents more complete and more detailed results.
We emphasize that there are no freely adjustable parameters in our approach.

We made use of these approximations to
calculate analytical laws for the bubble's collapse, afterbounce behaviour,
expansion dynamics, maximum radius, and expansion ratio.
With these results we could
clarify parameter dependences of numerically calculated curves
of diffusive equilibria in $P_a$--$R_0$ parameter space as those in
figure~\ref{total}. A summary of relevant analytical relations and
predictions for experimental verification was already given in the
Introduction.

An approximation of the RP equation by a Mathieu equation has
explained the wiggly resonance structure characteristic for many quantities
derived from RP dynamics. The concept of a
quasistatic Blake threshold between regimes of weakly oscillating and strongly
collapsing bubbles was able to shed light on the existence of
{\em stable} diffusive equilibria in the SL regime for high driving pressures. 
The change of sign in the
slope of $\left< p_{gas}\right>_{4}(R_0)$ is a generic feature of RP
dynamics, resulting from the dominance of surface tension pressure at small
$R_0$. This allows the bubble to reach a stable diffusive equilibrium.

In all approximations of the RP equation, the fluid viscosity term for
water or similar liquids
could be neglected without causing large errors. 
Both numerical computations and analytical estimates
of the magnitude of $p_{vis}$ show that $\eta_l$ has to be as high as
$\sim 40\,\eta_{\em water}$ to become a dominant contribution to bubble
dynamics.
Viscosity does, however, have a strong influence on the dynamics of surface
oscillations; parametric instabilities are weakened for larger $\eta_l$. 

Surface tension is the underlying cause for the
change from unstable to stable diffusive equilibria, which
stabilizes small bubbles to an extent that they can only 
show weak oscillations.
For fluids with low $\sigma$, collapses of bubbles of a given size are more
violent. This is especially interesting if this effect is combined with
higher
fluid viscosity to establish bubbles which show a similarly violent
collapse dynamics as bubbles in water while maintaining larger radii.

Other possibilities for an upscaling of the collapse intensity are
the use of lower driving frequencies $\om$ or of larger ambient pressures
$P_0$ at the same $P_a/P_0$. These predictions offer a useful guideline to
experimenters in search of upscaled single bubble sonoluminescence.

\begin{acknowledgments}
This work has been supported by the DFG uder grant SFB\,185-D8 and by the
joint DAAD/NSF Program for International Scientific Exchange. 
\end{acknowledgments}

\begin{appendix}

\section{Modifications of the RP ODE}
\label{subsecmod}
As stated in the Introduction, a lot of variations to the RP equation
(\ref{rp}) are known
from literature, see Lastman \& Wentzell (1981,\,1982)
for an overview. We mention here
the form derived by Flynn (1975a,\,1975b)
\begin{eqnarray}
\rho_l \left[ \left(1-{\dot{R}\over c_l}\right) R \ddot R + {3\over 2}
\left(1-{\dot{R}\over 3c_l}\right) \dot R^2 \right]  &=&
\left(1+{\dot{R}\over c_l}\right)\left[p_{gas}(R,t) - P(t) - P_0 - 4 \eta_l 
{\dot R \over R} - {2\sigma \over R}\right]
             \nonumber \\
&+& \left(1-{\dot{R}\over c_l}\right){R\over c_l} {{\mbox d}\over
{\mbox d}t}
\left[p_{gas}(R,t) - 4 \eta_l {\dot R \over R} - {2\sigma \over R}
\right],
\label{flynn}
\end{eqnarray}
which contains correction terms of higher order in $\dot{R}/c_l$.
It also includes
time derivatives of the surface tension and viscosity terms.

Gilmore's equation (see e.g.\ Hickling 1963) differs from the other RP
variations in
that its key variable is not pressure, but the enthalpy $H$ of the gas
at the bubble wall:
\begineq
\rho_l \left[ \left(1-{\dot{R}\over C_l}\right) R \ddot R +
{3\over 2} \left(1-{\dot{R}\over 3C_l}\right) \dot R^2 \right]  =
\left(1+{\dot{R}\over C_l}\right) H +
 \left(1-{\dot{R}\over C_l}\right){R\over C_l} {{\mbox d}\over
 {\mbox d}t} H.
\label{gilmore}
\endeq
Here, the sound speed $C_l$ is not a constant, but depends on $H$. The exact
form of this dependence has to be specified by an equation of state for
water, e.g.\ of modified Tait form (Prosperetti \& Lezzi 1986, Cramer 1980).
Gilmore's equation was shown in \cite{hic63}
to be an accurate description of a collapsing
{\em cavity} up to Mach numbers $|M_l|=|\dot{R}|/C_l$ as high as 5.
Its validity for the present problem of collapsing gas bubbles is, however,
not well established (Prosperetti 1984).

Figure \ref{roft}($a$) compares the bubble radius dynamics $R(t)$
computed from Eqs. (\ref{flynn}) and (\ref{gilmore}) to the solution of
(\ref{rp}). On this scale, the curves are almost indistinguishable.
Only a blow-up of the
region around the radius minimum reveals deviations.

As all equations (\ref{rp}),\,(\ref{flynn}),\,(\ref{gilmore}) have a common
limit
for small $M_l$, significant deviations are
only to be expected
during collapses, when the bubble wall velocity becomes of order of the
sound speed. Figure \ref{roft}($c$) shows $M_l$ for the
different $R(t)$ dynamics. Obviously, large differences occur only around
the main collapse, and they are not important for the overall dynamics
of the bubble, the collapse time interval being exceedingly small.
Thus, the RP equation provides a relatively simple and very accurate
description of bubble wall motion, which is confirmed by comparison with
recent experimental measurements of $R(t)$ in Tian \etal (1996). Note also that
our criterion (\ref{mach}) for energy transfer uses $M_{g}$, the Mach
number with respect to $c_{g}$, which (for argon) is
almost 5 times smaller than $c_l$ (in water). Therefore, the energy transfer
threshold can be well computed within the RP-SL approach.

\section{Length of afterbounce interval}\label{secablength}

The Mathieu model equation (\ref{mathieusurf}) can only be expected to
give an accurate
description of RP dynamics in the time interval $[\pi/2,3\pi/2]$,
for which it was matched to the Hill equation (\ref{hill}) via (\ref{equal}).
Therefore, we have to compare the afterbounce interval of RP dynamics to
this constant interval of length $\pi$.
Figure~\ref{dynmathrp} shows that for large driving $p$ the Mathieu dynamics
gives a good approximation for the end point of the afterbounce interval
(starting point of expansion phase), but fails to model the onset time of
afterbounces, i.e., the dimensionless collapse time $x^*$.
Thus, the length of the afterbounce interval is smaller than $\pi$ by a factor
of $C(p)=(3\pi/2-x^*(p))/\pi$.

The collapse time $x^*$ is relatively close to the maximum time $x_m$. It
is therefore convenient to compute it from an expansion of
(\ref{expright}$b$) in powers of $(x-\pi/2)$ and $(p-\pi/2)$. This yields
\begineq
x^*\approx 2.55+1.72 (p-\pi/2) - 1.31 (p-\pi/2)^2\, .
\label{xstar}
\endeq
The coefficients can be computed analytically, but are of very complicated
form. This expression corresponds to a correction factor for the length of
the afterbounce
interval and, equivalently, for the resonance order $k$, of
\begineq
C\approx 0.688 -0.548 (p-\pi/2) + 0.418 (p-\pi/2)^2\, .
\label{ckexp}
\endeq
This is the correction introduced in Section
\ref{secafter} which is important for a satisfactory description of the
resonances of (\ref{rp}) by (\ref{mathieusurf}), see figure~\ref{comprpmath}.

\end{appendix}

%\bibliography{sl_literatur}

\end{document}